\documentclass[a4paper,11pt]{article}

\usepackage{jheppub} 

\usepackage[T1]{fontenc} 
\usepackage{mathrsfs}

\title{\boldmath  Non-Abelian  chiral kinetic theory with self-consistent semiclassical expansion}

\author[a]{Xiao-Li Luo}
\author[a]{Shu-Xiang Ma}
\author[a,b,1]{Jian-Hua Gao.\note{Corresponding author.}}

\affiliation[a]{Shandong Provincial Key Laboratory of Nuclear Science, Nuclear Energy Technology and Comprehensive Utilization,
Weihai Frontier Innovation Institute of Nuclear Technology, School of Nuclear Science, Energy and Power Engineering, Shandong University, Shandong 250061, China}

\affiliation[b]{Weihai Research Institute of Industrial Technology of Shandong University, Weihai
264209, China}

\emailAdd{xiaoli\_luo@mail.sdu.edu.cn}
\emailAdd{shuxiangma@mail.sdu.edu.cn}
\emailAdd{gaojh@sdu.edu.cn}

\abstract{We derive the chiral kinetic theory in a non-Abelian gauge field using a self-consistent semiclassical expansion.
Within this new expansion scheme, we disentangle the Wigner equations up to second order and demonstrate that they do not introduce additional constraint equations.
By integrating the covariant chiral kinetic equations in eight-dimensional phase space, we obtain the corresponding equations in seven-dimensional phase space. This reduction facilitates numerical simulations and practical applications.

 }

\begin{document}
\maketitle
\flushbottom

\section{Introduction}
\label{sec:intro}

The spin degree of freedom serves as s a unique probe for studying the hot and dense matter created in high-energy heavy-ion collisions\cite{Liang:2004ph,Liang:2004xn,Gao:2019znl}.
At high temperatures where fermion masses become negligible, the chiral kinetic theory (CKT) provides a robust framework for
incorporating spin and non-equilibrium effects, leading to significant advancements in relativistic heavy-ion collision
\cite{Gao:2012ix,Stephanov:2012ki,Son:2012zy,Manuel:2013zaa,Chen:2012ca,Chen:2014cla,Chen:2015gta,Hidaka:2016yjf,Mueller:2017arw,Gorbar:2017cwv,
Huang:2018wdl,Gao:2018wmr,Carignano:2018gqt,Liu:2018xip,Carignano:2019zsh,Lin:2019fqo,Gao:2019zhk,Hayata:2020sqz,Luo:2021uog,Mameda:2023ueq,Yamamoto:2023okm}.
Recent reviews on the development of the  CKT can be found in Refs.\cite{Liu:2020ymh,Gao:2020pfu,Hidaka:2022dmn}.
The chiral kinetic equation (CKE) naturally emerges from the Wigner function approach in quantum gauge field theory
\cite{Hidaka:2016yjf,Huang:2018wdl,Gao:2018wmr,Liu:2018xip,Lin:2019fqo,Luo:2021uog}.
However, extending this approach from Abelian to non-Abelian gauge fields reveals a critical distinction: certain Wigner equations yield extra constraint equations
for non-Abelian fields that are automatically satisfied in the Abelian case\cite{Ochs:1998qj,Luo:2021uog}.
These extra constraints in non-Abelian gauge field originate from the fact that
the ``covariant gradient expansion'' used in \cite{Ochs:1998qj,Luo:2021uog} does not fully align with  the semiclassical $\hbar$ expansion
for non-Abelian gauge field.

In this work, we present a systematic derivation of the non-Abelian CKE  up to  second order using the  Wigner function approach,
where  a self-consistent semiclassical expansion is established through rescaling the gauge potential or the field strength tensor by $\hbar$.
This approach eliminates the extra constraint equations and enables consistent order-by-order solutions.

The paper is organized as follows:
Sec.\ref{sec:quantum} reviews   essential results from Refs.\cite{Ochs:1998qj,Luo:2021uog};
Sec.\ref{sec:constraint} analyzes the origin of the constraint equations   in non-Abelian Wigner equations  within the ``covariant gradient expansion'';
Sec.\ref{sec:disentangling} introduces our new expansion method and derives the  covariant CKE  to second order;
Sec.\ref{sec:ckt-7} integrates  the  eight-dimensional phase space formulation to seven dimensions;
Finally, Sec.\ref{sec:summary} summarizes this work.

Throughout this work, we use the  metric $g^{\mu\nu}=\mathrm{diag}(1,-1,-1,-1)$,
Levi-Civita tensor $\epsilon^{0123}=1$ and  natural units $\hbar=c=1$, explicitly restoring $\hbar$  only needed to clarify the
quantum corrections.

\section{Non-Abelian wigner functions and equations}
\label{sec:quantum}

In quantum gauge field theory, the Wigner functions are defined as  \cite{Heinz:1983nx,Elze:1986qd}
\begin{eqnarray}
\label{wigner-hat}
 W(x,p)=\int\frac{d^4 y}{(2\pi)^4} e^{-ip\cdot y}\left \langle  \rho \left(x+\frac{y}{2},x-\frac{y}{2}\right)\right \rangle,
\end{eqnarray}
where $\langle \cdots \rangle$ denotes the ensemble average. The  density operator $ \rho $ is a  gauge-covariant matrix operator  with both spinor and color indices
\begin{eqnarray}
\label{density-element}
\rho_{\alpha\beta}^{ij} \left(x+\frac{y}{2},x-\frac{y}{2}\right) = \bar\psi_\beta^{j'}\left(x+\frac{y}{2}\right)
U^{j'j}\left(x+\frac{y}{2},x\right) U^{ii'}\left(x,x-\frac{y}{2}\right) \psi^{i'}_{\alpha}\left(x-\frac{y}{2}\right),
\end{eqnarray}
where $\alpha,\beta$ denote spinor components of Dirac field $\psi$ or $ \bar\psi$ while $i,i',j,j'$ denote color components in fundamental representation of gauge group.
 The gauge link or wilson line $U^{ij}$ is defined by
\begin{eqnarray}
\label{link}
U^{ij}(x,y)=\left[{P}\exp\left(ig\int_y^x dz^\mu  A_\mu(z)\right)\right]^{ij}
\end{eqnarray}
where  $P$ denotes  path ordering of the operator  along the straight path from $x$ to $y$.
The gauge field potential is defined by $A_\mu=A^a_\mu t^a$, The $N^2-1$ hermitian generators $t^a$ of $SU(N)$ in the fundamental representation satisfying
\begin{eqnarray}
\label{Lie}
\textrm{Tr}\, t^a =0,\ \ \
\left[t^a,t^b\right]= if^{abc}t^c,\ \ \
\left\{t^a,t^b\right\} = \frac{1}{N}\delta^{ab}{\bf 1}+ d^{abc}t^c.
\end{eqnarray}
Within this basis, the Wigner function can be decomposed into
\begin{eqnarray}
\label{W-I-a}
W(x,p)&=&W^{I}(x,p) {\bf{1}} + W^{a}(x,p) t^a,
\end{eqnarray}
where $W^{I}(x,p)$ and $W^{a}(x,p)$ denote the color-singlet and color-multiplet, respectively, and  can be obtained by computing the trace in color space
\begin{eqnarray}
W^{I}(x,p) &=& \frac{1}{N} \textrm{tr} W(x,p), \, \ \ \
W^{a}(x,p)   = { 2} \textrm{tr}\left[ t^a W(x,p) \right].
\end{eqnarray}

The covariant derivative in  fundamental representation is defined as
\begin{eqnarray}
\label{Dmu}
D_\mu(x)=\partial_\mu -ig A_\mu(x),\ \ \ D^\dagger_\mu(x)=\partial_\mu^\dagger + ig A_\mu(x),
\end{eqnarray}
where $\partial_\mu$ and $\partial_\mu^\dagger$ act on the right and the left, respectively.
It follows that  the field strength tensor is given by
\begin{eqnarray}
\label{Fmunu}
F_{\mu\nu}(x)&\equiv& F_{\mu\nu}^a t^a =-\frac{1}{ig} \left[ D_\mu,D_\nu\right] = \partial_\mu A_\nu(x)-\partial_\nu A_\mu(x)-ig\left[A_\mu(x),A_\nu(x)\right].
\end{eqnarray}
In this paper, we will consider  massless fermions and treat  non-Abelian gauge field as a background field. Under such approximation,  the Wigner equation satisfied by Wigner function read \cite{Ochs:1998qj,Luo:2021uog}
\begin{eqnarray}
\label{equation-5}
\gamma^\mu \left(p_\mu +\frac{1}{2}i\mathscr{D}_\mu(x)\right) W
&=&-\frac{i g}{2}\gamma^\mu \partial^\nu_p \left\{  \int_0^1 d s  {\,} \frac{1+s}{2}\left[ e^{-\frac{1}{2} i s\Delta}  F_{\mu\nu}(x)\right]  W\right.\nonumber\\
& &\left. \hspace{1.5cm}+  W\int_0^1 d s  {\,} \frac{1-s}{2}
\left[ e^{\frac{1}{2}is\Delta} F_{\mu\nu}(x) \right]\right\},\\
\label{equation-6}
 W\gamma^\mu \left(p_\mu -\frac{1}{2}i\mathscr{D}^\dagger_\mu(x)\right)
&=&\frac{i g}{2} \partial^\nu_p \left\{  \int_0^1 d s  {\,} \frac{1-s}{2}
\left[ e^{-\frac{1}{2}is\Delta} F_{\mu\nu}(x) \right]  W\right.\nonumber\\
& &\left.\hspace{1.5cm} +  W\int_0^1 d s  {\,} \frac{1+s}{2}\left[ e^{\frac{1}{2} i s\Delta}  F_{\mu\nu}(x)\right]\right\}
\gamma^\mu,
\end{eqnarray}
where the second equation can be obtained from the  hermitian adjoint  of the first one.
In the Wigner equations above, we have  introduced the  covariant derivative
\begin{eqnarray}
\mathscr{D}_\mu (x) W \equiv \left[{D}_\mu (x), W \right]=\partial_\mu^x W - ig \left[ A_\mu(x), W\right],
\end{eqnarray}
and the  conjugate
\begin{eqnarray}
 W \mathscr{D}^\dagger_\mu (x) \equiv \left[W ,{D}^\dagger_\mu (x)\right]=\partial_\mu^x W - ig \left[ A_\mu(x), W\right],
\end{eqnarray}
in the adjoint representation and the triangle operator  $\Delta\equiv \partial_p \cdot \mathscr{D}(x)$  with  $\mathscr{D}(x)$ only acting on $ F_{\mu\nu}$ and $\partial_p$ always on $ W$ after or before it. With the convention in \cite{Ochs:1998qj}
\begin{eqnarray}
\label{convention}
 W \partial^{\nu_1}_p \cdots  \partial^{\nu_k}_p & \equiv & (-1)^k  \partial^{\nu_k}_p \cdots  \partial^{\nu_1}_p  W,
\end{eqnarray}
and the  definitions for generalized non-local momentum and derivative operators
\begin{eqnarray}
\label{Pi}
\Pi_\mu &=& p_\mu  + \frac{g}{2}\int_0^1 d s  {\,} \left( e^{-\frac{1}{2} i s\Delta}  F_{\mu\nu}(x)\right)  i s \partial^\nu_p, \\
G_\mu &=& D_\mu  + \frac{g}{2}\int_0^1 d s  {\,} \left( e^{-\frac{1}{2} i s\Delta}  F_{\mu\nu}(x)\right)    \partial^\nu_p,
\end{eqnarray}
we can recast  the Wigner equations into a more compact form \cite{Ochs:1998qj},
\begin{eqnarray}
\label{equation-LR1}
0 &=& \left\{\gamma^\mu,\left\{\Pi_\mu,   W(x,p)\right\}\right\}
 + i \left[ \gamma^\mu, \left[G_\mu,   W(x,p)\right]\right],\\
\label{equation-LR2}
0 &=& \left[\gamma^\mu,\left\{\Pi_\mu,   W(x,p)\right\}\right]
 + i \left\{ \gamma^\mu, \left[G_\mu,   W(x,p)\right]\right\}.
\end{eqnarray}
After decomposing the Wigner function in the spinor space
\begin{eqnarray}
\label{decomposition}
  W=\frac{1}{4}\left[{ {\mathscr{F}}}+i\gamma^5{ {\mathscr{P}}}+\gamma^\mu { { \mathscr{V}}}_\mu +\gamma^\mu\gamma^5{  {\mathscr{A}}}_\mu
+\frac{1}{2}\sigma^{\mu\nu} { {\mathscr{S}}}_{\mu\nu}\right],
\end{eqnarray}
and defining  the chiral Wigner function
\begin{eqnarray}
\label{Js}
 {\mathscr{J}}^\mu_s &=& \frac{1}{2}\left( {\mathscr{V}}^\mu + s {\mathscr{A}}^\mu\right),
\end{eqnarray}
 with $s =+1/-1$  right-handed/left-handed chirality,  the equations for the chiral Wigner function ${\mathscr{J}}^\mu_s$
 will decouple from all the other components:
\begin{eqnarray}
\label{Js-c1}
0 &=& \left\{\Pi_\mu,  {\mathscr{J}}^\mu_s \right\},\\
\label{Js-t}
0 &=& \left[G_\mu,  {\mathscr{J}}^\mu_s \right],\\
\label{Js-c2}
0&=& \left\{\Pi^\mu,  {\mathscr{J}}^{\nu}_s \right\} - \left\{\Pi^\nu,  {\mathscr{J}}^{\mu}_s \right\}
 {+} s \epsilon^{\mu\nu\alpha\beta} \left[G_\alpha,  {\mathscr{J}}_{s\beta} \right].
\end{eqnarray}
These Wigner equations will be  the starting point for our current work.

\section{Constraint equations in the covariant gradient expansion }
\label{sec:constraint}

In this section, we will review how the constraint equations emerge within the covariant gradient expansion or triangle expansion
used in Refs. \cite{Elze:1986qd,Ochs:1998qj,Luo:2021uog}. The covariant gradient expansion denotes the expansion with ${D}_\mu$ or  ${\mathscr{D}}_\mu$  and preserves gauge covariant order by order
in the Wigner equations (\ref{Js-c1})-(\ref{Js-c2}). Let us  recover the dependence
of $\hbar $ for the Wigner equations
\begin{eqnarray}
\label{Js-c1-h}
0 &=& \left\{\Pi_\mu,  {\mathscr{J}}^\mu_s \right\},\\
\label{Js-t-h}
0 &=& \left[\hbar G_\mu,  {\mathscr{J}}^\mu_s \right],\\
\label{Js-c2-h}
0&=& \left\{\Pi^\mu,  {\mathscr{J}}^{\nu}_s \right\} - \left\{\Pi^\nu,  {\mathscr{J}}^{\mu}_s \right\}
 {+} s \epsilon^{\mu\nu\alpha\beta} \left[\hbar G_\alpha,  {\mathscr{J}}_{s\beta} \right],
\end{eqnarray}
and  the generalized non-local momentum and derivative operators
\begin{eqnarray}
\label{Pi-1}
\Pi_\mu &=& \sum_{k=0}^\infty \hbar^k \Pi_\mu^{(k)} = p_\mu  - \frac{i  \hbar g}{2 } \sum_{k=0}^\infty \left(-\frac{i}{2}\right)^k  \frac{k+1}{(k+2)!}
\left[\left(\hbar\partial_p\cdot  {\mathscr{D}} \right)^k F_{\nu\mu}\right] \partial_p^\nu \\
\label{G-1}
G_\mu &=&\sum_{k=0}^\infty \hbar^{k} G_\mu^{(k)} = D_\mu  - \frac{ g}{2 }\sum_{k=0}^\infty \left(-\frac{i}{2}\right)^k  \frac{1}{(k+1)!}
\left[\left(\hbar \partial_p\cdot  {\mathscr{D}} \right)^k F_{\nu\mu} \right] \partial_p^\nu.
\end{eqnarray}
We list these operators up to  second order within covariant gradient expansion
\begin{eqnarray}
\label{Pi-012}
\Pi_\mu^{(0)} &=& p_\mu,\ \ \
\Pi_\mu^{(1)} = \frac{i g}{4} F_{\mu\nu}\partial^\nu_p,\ \
\Pi_\mu^{(2)} = \frac{ g}{12 }\left[\left(\partial_p\cdot  {\mathscr{D}} \right) F_{\mu\nu} \right] \partial_p^\nu,\\
\label{G-01}
G_\mu^{(0)} &=&  D_\mu  +\frac{g}{2} F_{\mu\nu}\partial^\nu_p, \ \
G_\mu^{(1)} = - \frac{i g}{8}\left[\left(\partial_p\cdot  {\mathscr{D}} \right) F_{\mu\nu} \right] \partial_p^\nu.
\end{eqnarray}
We note  that the covariant gradient  always appears along with $\hbar$ and seems as if the covariant gradient expansion  is identical as
the semiclassical expansion of $\hbar $. Actually, such expansion is not completely identical to the expansion in powers of $\hbar$ \cite{Elze:1986qd,Ochs:1998qj}  for non-Abelian gauge field though it is identical for Abelian gauge field. Such inconsistency leads to the extra constraint equations for non-Abelian Wigner equations
which are absent for Abelian Wigner equations.  Now we can show how the constraint equations emerge for  non-Abelian Wigner equations.
To do it, we assume that the Wigner functions can be expanded as
\begin{eqnarray}
  W (x,p) =\sum_{k=0}^\infty \hbar^k   W^{(k)}(x,p),
\end{eqnarray}
and introduce  time-like constant 4-vector $n^\mu$ with normalization $n^2=1$.
Then we  can  decompose any vector $X^\mu$ into time-like and space-like components
\begin{equation}
X^\mu=X_n n^\mu + \bar X^\mu,
\end{equation}
where $X_n=X\cdot n$ and $\bar X^\mu = \Delta^{\mu\nu}X_\nu$ with $\Delta^{\mu\nu}=g^{\mu\nu}-n^\mu n^\nu$.
The Wigner functions and  equations can be also decomposed along the direction $n^\mu$ order by order.
The  zeroth-order result reads
\begin{eqnarray}
\label{J-c1-n-0-W}
0 &=& p_n {\mathscr{J}}_n^{(0)} +\bar p_\mu \bar{{\mathscr{J}}}^{(0)\mu},\\
\label{J-t-n-0-W}
0 &=& \left[G_n^{(0)},{\mathscr{J}}_n^{(0)}\right]+\left[\bar G_\mu^{(0)}, \bar{{\mathscr{J}}}^{(0)\mu} \right],\\
\label{J-c2-n-0-W}
0&=& \bar p ^\mu {\mathscr{J}}_n^{(0)}  - p_n \bar{{\mathscr{J}}}^{(0)\mu},\\
\label{J-c2-bar-0-W}
0&=& \bar p^\mu \bar{{\mathscr{J}}}^{(0)\nu} - \bar p^\nu \bar{ {\mathscr{J}}}^{(0)\mu}.
\end{eqnarray}
From Eq.(\ref{J-c2-n-0-W}), we can express the space-like component $\bar{{\mathscr{J}}}^{(0)\mu}$ in terms of ${\mathscr{J}}_n^{(0)}$
\begin{eqnarray}
\label{Jnbar-Jn-0}
\bar{{\mathscr{J}}}^{(0)\mu} &=& \bar p ^\mu   \frac{{\mathscr{J}}_n^{(0)} }{p_n}.
\end{eqnarray}
Substituting Eq.(\ref{Jnbar-Jn-0}) into  Eq.(\ref{J-c2-bar-0-W}), we find  it  is automatically satisfied.

The first-order equations are given by
\begin{eqnarray}
\label{J-c1-n-1-W}
0 &=& 2 p_n{\mathscr{J}}^{(1)}_n +2  \bar p_\mu \bar{{\mathscr{J}}}^{(1)\mu}
+\left\{\Pi_n^{(1)}, {\mathscr{J}}^{(0)}_n \right\} + \left\{\bar\Pi_\mu^{(1)}, \bar{{\mathscr{J}}}^{(0)\mu} \right\},\\
\label{J-t-n-1-W}
0 &=& \left[G_n^{(0)}, {\mathscr{J}}^{{(1)}}_n \right]+\left[\bar G_\mu^{(0)}, \bar{{\mathscr{J}}}^{{(1)}\mu} \right]
+\left[G_n^{(1)}, {\mathscr{J}}^{{(0)}}_n \right]+\left[\bar G_\mu^{(1)},\bar{ {\mathscr{J}}}^{{(0)}\mu} \right],\hspace{8pt}\\
\label{J-c2-n-1-W}
0 &=& 2\bar p^\mu {\mathscr{J}}^{(1)}_n - 2 p_n \bar{{\mathscr{J}}}^{(1)\mu}
 + \left\{\bar\Pi^{(1)\mu},{\mathscr{J}}^{(0)}_n \right\}   -\left\{\Pi^{(1)}_n, \bar{{\mathscr{J}}}^{(0)\mu} \right\}\nonumber\\
& &  {+}s \bar\epsilon^{\mu\alpha\beta} \left[G_\alpha^{(0)}, {\mathscr{J}}_\beta^{(0)} \right],\\
 \label{J-c2-bar-1-W}
0 &=&2\bar p^{\mu} \bar{{\mathscr{J}}}^{(1)\nu}
 -2\bar p^{\nu} \bar{{\mathscr{J}}}^{(1)\mu}
 + \left\{\bar\Pi^{(1)\mu},\bar{ {\mathscr{J}}}^{(0)\nu} \right\}
 -\left\{\bar\Pi^{(1)\nu}, \bar{{\mathscr{J}}}^{(0)\mu} \right\}\nonumber\\
& & {+} s\bar \epsilon^{\mu\nu\alpha}\left( \left[G_\alpha^{(0)},{\mathscr{J}}_n^{(0)} \right]
-\left[G_n^{(0)}, {\mathscr{J}}_\alpha^{(0)} \right]\right).
\end{eqnarray}
From Eq.(\ref{J-c2-n-1-W}), we can express $\bar{{\mathscr{J}}}^{(1)\mu}$ in terms of ${\mathscr{J}}^{(1)}_n$ and ${\mathscr{J}}^{(0)}_n$
\begin{eqnarray}
\label{Jnbar-Jn-1a}
\bar{{\mathscr{J}}}^{(1)\mu} &=&\bar p^\mu\frac{ {\mathscr{J}}^{(1)}_n}{p_n}
 {+}\frac{s}{2 p_n}\bar\epsilon^{\mu\alpha\beta} \left[G_\alpha^{(0)}, \bar p_\beta\frac{ {\mathscr{J}}_n^{(0)}}{p_n} \right]\nonumber\\
& &+ \frac{1}{2 p_n}\left(\left\{\bar\Pi^{(1)\mu}, p_n \frac{{\mathscr{J}}^{(0)}_n}{p_n} \right\}
-  \left\{\Pi^{(1)}_n,  \bar p^\mu \frac{{\mathscr{J}}_n^{(0)}}{p_n} \right\}\right).
\end{eqnarray}
Plugging it into Eq.(\ref{J-c2-bar-1-W}) and using the zeroth-order results (\ref{J-c1-n-0-W}), (\ref{J-t-n-0-W}), and (\ref{Jnbar-Jn-0}),
we obtain the constraint equation
\begin{eqnarray}
\label{constraint-n-Jmu0}
0&=&\frac{ig}{2p_n}n_\alpha\left( \left[F^{\nu\alpha} ,  {\mathscr{J}}^{(0)\mu}\right]
+\left[F^{\alpha\mu} ,  {\mathscr{J}}^{(0)\nu} \right]
+\left[F^{\mu\nu} ,  {\mathscr{J}}^{(0)\alpha} \right]\right).
\end{eqnarray}
To show where this constraint equation originates , we need  the second-order result.
The second order expression of Eq.(\ref{Js-c2}) is given by
\begin{eqnarray}
\label{J-c2-n-2-W}
 0 &=& 2 \bar p^\mu {\mathscr{J}}^{(2)}_n - 2p_n\bar{ {\mathscr{J}}}^{(2)\mu}\nonumber\\
& & + \left\{\bar\Pi^{(1)\mu}, {\mathscr{J}}^{(1)}_n \right\}    -\left\{\Pi^{(1)}_n, \bar{{\mathscr{J}}}^{(1)\mu} \right\}
 + \left\{\bar\Pi^{(2)\mu}, {\mathscr{J}}^{(0)}_n \right\}    -\left\{\Pi^{(2)}_n, \bar{{\mathscr{J}}}^{(0)\mu} \right\}\nonumber\\
& & {+} s\bar \epsilon^{\mu\alpha\beta}\left( \left[G_\alpha^{(0)}, {\mathscr{J}}_\beta^{(1)} \right]
+\left[G_\alpha^{(1)}, {\mathscr{J}}_\beta^{(0)} \right]\right),\\
\label{J-c2-bar-2-W}
0&=& 2 \bar p^\mu \bar{{\mathscr{J}}}^{(2)\nu} - 2 \bar p^\nu \bar{ {\mathscr{J}}}^{(2)\mu}\nonumber\\
& & + \left\{\bar\Pi^{(1)\mu}, \bar{{\mathscr{J}}}^{(1)\nu} \right\}    - \left\{\bar\Pi^{(1)\nu},\bar{ {\mathscr{J}}}^{(1)\mu} \right\}
 + \left\{\bar\Pi^{(2)\mu},\bar{ {\mathscr{J}}}^{(0)\nu} \right\}  - \left\{\bar\Pi^{(2)\nu}, \bar{{\mathscr{J}}}^{(0)\mu} \right\}\nonumber\\
& &  {+} s \bar\epsilon^{\mu\nu\alpha}\left( \left[G_\alpha^{(0)}, {\mathscr{J}}_n^{(1)} \right]
- \left[G_n^{(0)}, {\mathscr{J}}_\alpha^{(1)} \right]
+\left[G_\alpha^{(1)}, {\mathscr{J}}_n^{(0)} \right]
-\left[G_n^{(1)}, {\mathscr{J}}_\alpha^{(0)} \right]\right).
\end{eqnarray}
The first equation above expresses $\bar{ {\mathscr{J}}}^{(2)\mu}$ in terms of ${\mathscr{J}}^{(2)}_n$, ${\mathscr{J}}^{(1)}_n$ and ${\mathscr{J}}^{(0)}_n$ as
\begin{eqnarray}
\label{Jnbar-Jn-1}
\bar{ {\mathscr{J}}}^{(2)\mu}&=&\bar p^\mu \frac{{\mathscr{J}}^{(2)}_n}{p_n}
 {+}\frac{s}{2p_n} \bar\epsilon^{\mu\alpha\beta}\left(\left[G_\alpha^{(0)}, {\mathscr{J}}_\beta^{(1)} \right]
 {+}\left[G_\alpha^{(1)}, {\mathscr{J}}_\beta^{(0)} \right]\right)\nonumber\\
& & + \frac{1}{2p_n}\left(\left\{\bar\Pi^{(1)\mu}, {\mathscr{J}}^{(1)}_n \right\}
 -\left\{\Pi^{(1)}_n, \bar{{\mathscr{J}}}^{(1)\mu} \right\} \right) \nonumber\\
& & + \frac{1}{2p_n}\left(\left\{\bar\Pi^{(2)\mu}, {\mathscr{J}}^{(0)}_n \right\}
  -\left\{\Pi^{(2)}_n, \bar{{\mathscr{J}}}^{(0)\mu} \right\}\right).
\end{eqnarray}
We note that   the second term of the right-handed side in  Eq.(\ref{Jnbar-Jn-1}) or the last term in Eq.(\ref{J-c2-bar-2-W})
can result in  the commutator $\left[ D_\mu, D_\nu \right] $ if we substitute the expression (\ref{Jnbar-Jn-1a}) into these terms.
When we recover the $\hbar$ dependence
\begin{eqnarray}
D_\mu(x)&=&\partial_\mu -\frac{ig}{\hbar} A_\mu(x),\ \ \ \left[ D_\mu, D_\nu \right] =- \frac{ig}{\hbar}F_{\mu\nu},
\end{eqnarray}
the commutator $[D_\mu,D_\nu]$ will contribute to  the lower order due to the $\hbar$ dependence in the denominator.
We find that all the terms from the commutator $[D_\mu,D_\nu]$ in Eq(\ref{J-c2-bar-2-W})  will  give rise to
exactly the same result as the right hand side in Eq.(\ref{constraint-n-Jmu0}) except for a minus sign. Hence,
the constraint equation (\ref{constraint-n-Jmu0}) originates from the order-mixing in the covariant gradient expansion.
Such issues does not exist for the Abelian case because only ordinary partial derivative $\partial_\mu$ appears in
Abelian Wigner equations.

\section{Wigner equations in self-consistent semiclassical expansion }
\label{sec:disentangling}

We can solve the issue associated with the order-mixing in covariant gradient expansion by rescaling the non-Abelian gauge potential
\begin{eqnarray}
\frac{1}{\hbar} A_\mu(x)  \rightarrow   A_\mu(x),
\end{eqnarray}
which denotes that the $1/\hbar$ factor has been absorbed into $ A_\mu(x) $. For brevity, we still use the same symbol  $A_\mu(x)$ to label the rescaled gauge potential. Then we have
\begin{eqnarray}
 \left[ D_\mu, D_\nu \right] =- ig F_{\mu\nu}.
\end{eqnarray}
There is no explicit factor  $1/\hbar$ any more because it has been absorbed into $F_{\mu\nu}$. After such rescaling, we can expand the operators as
\begin{eqnarray}
\label{Pi-1-new}
\Pi_\mu &=& \sum_{k=0}^\infty \hbar^k \Pi_\mu^{(k)} = p_\mu  - \frac{i  \hbar^2 g}{2 } \sum_{k=0}^\infty \left(-\frac{i}{2}\right)^k  \frac{k+1}{(k+2)!}
\left[\left(\hbar\partial_p\cdot  {\mathscr{D}} \right)^k F_{\nu\mu}\right] \partial_p^\nu \\
\label{G-1-new}
G_\mu &=&\sum_{k=0}^\infty \hbar^{k} G_\mu^{(k)} = D_\mu  - \frac{ g\hbar}{2 }\sum_{k=0}^\infty \left(-\frac{i}{2}\right)^k  \frac{1}{(k+1)!}
\left[\left(\hbar \partial_p\cdot  {\mathscr{D}} \right)^k F_{\nu\mu} \right] \partial_p^\nu.
\end{eqnarray}
Up to  second order within covariant gradient expansion, we have
\begin{eqnarray}
\label{Pi-012-new}
\Pi_\mu^{(0)} &=& p_\mu,\ \ \
\Pi_\mu^{(1)} = 0,\ \
\Pi_\mu^{(2)} = \frac{i g}{4} F_{\mu\nu}\partial^\nu_p,\\
\label{G-01-new}
G_\mu^{(0)} &=&  D_\mu , \ \
G_\mu^{(1)} =  \frac{g}{2} F_{\mu\nu}\partial^\nu_p, \ \
G_\mu^{(2)} = - \frac{i g}{8}\left[\left(\partial_p\cdot  {\mathscr{D}} \right) F_{\mu\nu} \right] \partial_p^\nu.
\end{eqnarray}
In contrast with the results in Eqs.(\ref{Pi-1})-(\ref{G-01}), we note that the terms associated with $F_{\mu\nu}$ become higher order by one due to the rescaling of
$ A_\mu(x) $. In the following, we will expand the Wigner functions and equations up to  second order within this new perturbative scheme.

\subsection{Zeroth-order result}
The zeroth-order Wigner equations retain the same expressions as in Eqs.(\ref{J-c1-n-0-W})-(\ref{J-c2-bar-0-W}) but with the new definitions for
$G_\mu^{(0)}$ as given  in Eq(\ref{G-01-new}).
Substituting Eq.(\ref{Jnbar-Jn-0}) into  Eq.(\ref{J-c1-n-0-W}), we have
\begin{eqnarray}
\frac{p^2}{p_n}\mathscr{J}^{(0)}_{n}&=&0,
\end{eqnarray}
which means that the Wigner function must be on-shell. The general solution is given by
\begin{eqnarray}
\label{Jn-0-delta}
\mathscr{J}^{(0)}_{n}&=& p_n  \mathcal{J}^{(0)}_{n} \delta\left(p^2\right),
\end{eqnarray}
where $\mathcal{J}^{(0)}_{n}$ is a non-singular distribution at $p^2$.
We can put the $\bar{\mathscr{J}}^{(0)\mu}$ and $\mathscr{J}^{(0)}_{n}$ together
\begin{eqnarray}
\label{Jmu-Jn-0}
{\mathscr{J}}^{(0)\mu} &=& p ^\mu   \frac{{\mathscr{J}}_n^{(0)} }{p_n}.
\end{eqnarray}
Substituting it into the Eq.(\ref{J-t-n-0-W}) gives rise to the zeroth-order covariant CKE in eight-dimensional phase space
\begin{eqnarray}
\label{cke-0-8d}
0 &=& \mathscr{D}_\mu {\mathscr{J}}^{(0)\mu}.
\end{eqnarray}
Obviously, the Eq.(\ref{J-c2-bar-0-W}) is automatically satisfied from the expression (\ref{Jnbar-Jn-0}).

The Wigner function ${\mathscr{J}}^{(0)\mu} $ is still a matrix in color space. After decomposing it in the light of the Eq.(\ref{W-I-a}), we have
\begin{eqnarray}
\label{Jmu-Jn-0-I}
{\mathscr{J}}^{(0)I}_\mu &=& p_\mu   \frac{{\mathscr{J}}_n^{(0)I} }{p_n} = p_\mu  \mathcal{J}^{(0)I}_{n} \delta\left(p^2\right),\\
\label{Jmu-Jn-0-a}
{\mathscr{J}}^{(0)a}_\mu &=& p_\mu   \frac{{\mathscr{J}}_n^{(0)a} }{p_n} = p_\mu  \mathcal{J}^{(0)a}_{n} \delta\left(p^2\right).
\end{eqnarray}
Correspondingly, the covariant CKE (\ref{cke-0-8d}) can be decomposed into
\begin{eqnarray}
\label{cke-0-8d-I}
0 &=& \partial^x_\mu  {\mathscr{J}}^{(0)I,\mu},\\
\label{cke-0-8d-a}
0 &=& {\mathscr{D}}^{ac}_\mu  {\mathscr{J}}^{(0)a,\mu},
\end{eqnarray}
where
\begin{eqnarray}
{\mathscr{D}}^{ac}_\mu &=& \delta^{ca}  \partial^x_\mu  + g  f^{bca} A^b_\mu.
\end{eqnarray}
Obviously, the color singlet and multiplet totally decouple with each other  at zeroth order within this new expansion scheme. Although the field strength tensor
$F_{\mu\nu}$ does not arise at zeroth order, the gauge potential $ A_\mu$ is still involved in the covariant derivation ${\mathscr{D}}^{ac}_\mu$.

\subsection{First-order result}

The first-order  Wigner equations in this new expansion scheme also retain the same form as given in  Eqs.(\ref{J-c1-n-1-W})-(\ref{J-c2-bar-1-W})
with  new assignment for $G_\mu^{(0)}$, $G_\mu^{(1)}$ and $\Pi_\mu^{(1)}$  in Eqs.(\ref{Pi-012-new}) and (\ref{G-01-new}).
We note that the operator  $\Pi_\mu^{(1)}$ vanishes with this new expansion.
Inserting Eq.(\ref{Jnbar-Jn-1a}) into (\ref{J-c1-n-0-W})  gives
\begin{eqnarray}
\frac{p^2}{p_n} {\mathscr{J}}^{(1)}_{n} &=&
 - \frac{s}{2p_n}\bar\epsilon_{\mu\alpha\beta}  \bar p^\mu \mathscr{D}^\alpha   {\mathscr{J}}^{(0)\beta}
 =0
\end{eqnarray}
which implies that the Wigner function is still on-shell at the first order in this new expansion. The general solution is given by
\begin{eqnarray}
\label{Jsn-onshell-1}
\mathscr{J}^{(1)}_{n}&=& p_n  \mathcal{J}^{(1)}_{n} \delta\left(p^2-m^2\right)
\end{eqnarray}
It is easy to verify that the Eq.(\ref{J-c2-bar-1-W}) holds automatically after we insert Eq.(\ref{Jnbar-Jn-1a})
into it and use the zeroth-order equation (\ref{cke-0-8d}).
Therefore, our new expansion scheme contains no constraint equations, which is why we characterize it as a self-consistent semiclassical expansion.

Putting $\bar{\mathscr{J}}^{(1)\mu}$ and $\mathscr{J}^{(1)}_{n}$ together, we have
\begin{eqnarray}
\label{Jmu-Jn-1}
{\mathscr{J}}^{(1)}_\mu &=& p_\mu   \frac{{\mathscr{J}}_n^{(1)} }{p_n}
+\frac{s}{2p_n}\bar\epsilon_{\mu\alpha\beta} \mathscr{D}^\alpha   {\mathscr{J}}^{(0)\beta}
\end{eqnarray}
Substituting it into the Eq.(\ref{J-t-n-1-W}) gives the covariant CKE in eight-dimensional phase space
\begin{eqnarray}
\label{cke-1-8d}
0 &=& \mathscr{D}_\mu {\mathscr{J}}^{(1)\mu}
+\frac{1}{2}  \partial^\nu_p\left\{ \mathscr{J}^{(0)\mu}, { F}_{\mu \nu} \right\}
\end{eqnarray}
After decomposing the Wigner functions in color space with the Eq.(\ref{W-I-a}), we have
\begin{eqnarray}
\label{Jmu-Jn-1-I}
{\mathscr{J}}^{(1)I}_\mu &=& p_\mu   \frac{{\mathscr{J}}_n^{(1)I} }{p_n}
+\frac{s}{2p_n}\bar\epsilon_{\mu\alpha\beta}\partial_x^\alpha   {\mathscr{J}}^{(0)I,\beta},\\
\label{Jmu-Jn-1-a}
{\mathscr{J}}^{(1)a}_\mu &=& p_\mu   \frac{{\mathscr{J}}_n^{(1)a} }{p_n}
+\frac{s}{2p_n}\bar\epsilon_{\mu\alpha\beta}\mathscr{D}^{\alpha,ac}   {\mathscr{J}}^{(0)c,\beta} .
\end{eqnarray}
The corresponding covariant CKEs for color singlet and multiplet are  given by
\begin{eqnarray}
\label{cke-1-8d-I}
0 &=& \partial^x_\mu  {\mathscr{J}}^{(1)I,\mu}
 +\frac{1}{2N} { F}_{\mu \nu}^a \partial^\nu_p  {\mathscr{J}}^{(0)a,\mu} \\
\label{cke-1-8d-a}
0 &=& {\mathscr{D}}^{ac}_\mu  {\mathscr{J}}^{(1)c,\mu} +  F_{\mu\nu}^a  \partial_p^\nu {\mathscr{J}}^{(0)I,\mu}
+ \frac{1}{2}d^{cba} { F}_{\mu \nu}^b \partial^\nu_p  {\mathscr{J}}^{(0)c,\mu}
\end{eqnarray}
We note that the constitutive equations for color singlet and multiplet of the  chiral Wigner functions in Eqs.(\ref{Jmu-Jn-1-I}) and (\ref{Jmu-Jn-1-a}) are
 independent on each other at  first order. There is no mixing between color singlet and multiplet.
 This  is similar as the results (\ref{Jmu-Jn-0-I}) and (\ref{Jmu-Jn-0-a}) at zeroth order. However, in the transport equations (\ref{cke-1-8d-I}) and (\ref{cke-1-8d-a}),
 the color singlet and multiplet components are coupled. Specifically, the first-order kinetic equation for the color singlet is coupled with the zero-order color multiplet,
 while the first-order kinetic equation for the color multiplet is coupled with the zero-order color singlet.

\subsection{Second-order result}
In self-consistent semiclassical expansion, there is no modification to the on-shell condition up to first order. To get the modification to  on-shell condition, we need to go beyond  first order.   The Wigner equations at second order are given by
\begin{eqnarray}
\label{F-2-st}
0 &=& p_\mu {\mathscr{J}}^{(2)\mu}
-\frac{i}{8} \partial^\nu_p \left[\mathscr{J}^{(0)\mu}, { F}_{\mu \nu}  \right],\\
\label{P-2-st}
0 &=& \mathscr{D}^\mu {{ \mathscr{J}}}_{\mu}^{(2)}
+\frac{1}{2} \partial^\nu_p \left\{\mathscr{J}^{(1)\mu}, { F}_{\mu \nu}  \right\}
+\frac{i}{8} \partial^\nu_p \partial^\lambda_p
 \left[\mathscr{J}^{(0)\mu},  (\mathscr{D}_\lambda { F}_{\mu \nu}) \right],\\
\label{S-2-t}
0 &=&\bar p_\mu {\mathscr{J}}^{(2)}_{n} - p_n \bar{\mathscr{J}}^{(2)}_{\mu}
+\frac{1}{2}s\bar\epsilon_{\mu\alpha\beta} \mathscr{D}^\alpha   {\mathscr{J}}^{(1)\beta}\nonumber\\
& &+\frac{1}{4}s\bar\epsilon_{\mu\alpha\beta} \partial^\lambda_p \left\{\mathscr{J}^{(0)\beta}, {{ F}^\alpha}_{\lambda} \right\}
+\frac{i}{8}\partial^\lambda_p \left(\left[\mathscr{J}^{(0)}_{\mu}, { F}_{n \lambda}\right]
- \left[\mathscr{J}^{(0)}_{n}, { F}_{\mu \lambda}\right]  \right),\\
\label{S-2-s}
0 &=&\bar p_\mu \bar{\mathscr{J}}^{(2)}_{\nu} - \bar p_\nu \bar{\mathscr{J}}^{(2)}_{\mu}
+\frac{1}{2}s\bar\epsilon_{\mu\nu\alpha}
\left( \mathscr{D}^\alpha   {\mathscr{J}}^{(1)}_{n} - \mathscr{D}_n    {\mathscr{J}}^{(1)\alpha}  \right)\nonumber\\
& &+\frac{1}{4}s\epsilon_{\bar\mu\bar\nu\alpha\beta} \partial^\lambda_p \left\{\mathscr{J}^{(0)\beta}, {{ F}^\alpha}_{\lambda} \right\}
+\frac{i}{8}\partial^\lambda_p \left(\left[\mathscr{J}^{(0)}_{\bar\mu}, { F}_{\bar\nu \lambda}\right]
- \left[\mathscr{J}^{(0)}_{\bar\nu}, { F}_{\bar\mu \lambda}\right]  \right)
\end{eqnarray}
From Eq.(\ref{S-2-t}), we can express $\bar{\mathscr{J}}^{(2)}_{\mu}$ in terms of $ {\mathscr{J}}^{(2)}_{n}$ and lower-order Wigner functions
\begin{eqnarray}
\label{Js-s-2}
\bar{\mathscr{J}}^{(2)}_{\mu}
&=& \frac{\bar p_\mu}{p_n} {\mathscr{J}}^{(2)}_{n}
+\frac{s}{2p_n}\bar\epsilon_{\mu\alpha\beta} \mathscr{D}^\alpha   {\mathscr{J}}^{(1)\beta}
+\frac{s}{4p_n}\bar\epsilon_{\mu\alpha\beta} \partial^\lambda_p \left\{\mathscr{J}^{(0)\beta}, {{ F}^\alpha}_{\lambda} \right\}\nonumber\\
& & +\frac{i}{8p_n}\partial^\lambda_p \left(\left[\mathscr{J}^{(0)}_{\mu}, { F}_{n \lambda}\right]
- \left[\mathscr{J}^{(0)}_{n}, { F}_{\mu \lambda}\right]  \right)
\end{eqnarray}
Inserting Eq.(\ref{Js-s-2}) into Eq.(\ref{F-2-st}),
we obtain
\begin{eqnarray}
\label{Jsn-onshell-2}
\frac{p^2}{p_n} {\mathscr{J}}^{(2)}_{n}&=& \frac{s}{4p_n} \bar\epsilon_{\mu\alpha\beta} \left\{\mathscr{J}^{(0)\beta}, {{ F}^{\alpha\mu}} \right\}
+ \frac{1}{4p_n^3}\left(\bar p^\mu \bar p^\nu - \bar p^2 \Delta^{\mu\nu}\right)
\bar{\mathscr{D}}_\mu \bar{\mathscr{D}}_\nu  {\mathscr{J}}^{(0)}_{n}\nonumber\\
& &
+ \frac{i}{4} \partial^\nu_p \left[\frac{p^\mu}{p_n}\mathscr{J}_{n}^{(0)}, { F}_{\mu \nu}  \right]
\end{eqnarray}
Non-vanishing of the right-hand-side of the equation implies the modification to the on-shell condition for free particles.  Given the zeroth-order results (\ref{Jn-0-delta}) and (\ref{Jmu-Jn-0}),
 we find that the general solution for ${\mathscr{J}}^{(2)}_{n}$  must take the form
\begin{eqnarray}
\label{Jsn-onshell-2}
\mathscr{J}^{(2)}_{n}&=& p_n  \mathcal{J}^{(2)}_{n} \delta\left(p^2\right)
-\frac{s}{4} \bar\epsilon_{\mu\alpha\beta}\bar p^\beta
 \left\{ \mathcal{J}^{(0)}_{n}, {{ F}^{\alpha\mu}} \right\} \delta'\left(p^2\right)\nonumber\\
& &- \frac{1}{4p_n}\left(\bar p^\mu \bar p^\nu -\bar p^2 \Delta^{\mu\nu}\right)
\bar{\mathscr{D}}_\mu \bar{\mathscr{D}}_\nu \mathcal{J}^{(0)}_{n} \delta'\left(p^2\right)
- \frac{i}{4}p_n \partial^\nu_p \left[p^\mu\mathcal{J}_{n}^{(0)}, { F}_{\mu \nu}  \right] \delta'\left(p^2\right)
\end{eqnarray}
Inserting Eq.(\ref{Js-s-2})  into Eq.(\ref{S-2-s}) and using the equations at zeroth and first order,
we verify that the equation is also satisfied automatically and there are no constraint equations at  second order.
We expect that the constraint equations will be absent  under this self-consistent semiclassical expansion up to any higher order.
It follows that the full second-order Wigner functions read
\begin{eqnarray}
\label{Jmu-Jn-2}
{\mathscr{J}}^{(2)\mu} &=& \frac{ p_\mu}{p_n} {\mathscr{J}}^{(2)}_{n}
+\frac{s}{2p_n}\bar\epsilon_{\mu\alpha\beta} \mathscr{D}^\alpha   {\mathscr{J}}^{(1)\beta}
+\frac{s}{4p_n}\bar\epsilon_{\mu\alpha\beta} \partial^\lambda_p \left\{\mathscr{J}^{(0)\beta}, {{ F}^\alpha}_{\lambda} \right\}\nonumber\\
& & +\frac{i}{8p_n}\partial^\lambda_p \left(\left[\mathscr{J}^{(0)}_{\mu}, { F}_{n \lambda}\right]
- \left[\mathscr{J}^{(0)}_{n}, { F}_{\mu \lambda}\right]  \right)
\end{eqnarray}
Plugging it into  Eq.(\ref{P-2-st}) leads to the covariant CKE in eight-dimensional phase space
\begin{eqnarray}
0 &=& \mathscr{D}^\mu {{ \mathscr{J}}}_{\mu}^{(2)}
+\frac{1}{2} \partial^\nu_p \left\{\mathscr{J}^{(1)\mu}, { F}_{\mu \nu}  \right\}
+\frac{i}{8} \partial^\nu_p \partial^\lambda_p
 \left[\mathscr{J}^{(0)\mu},  (\mathscr{D}_\lambda { F}_{\mu \nu}) \right]
\end{eqnarray}
After decomposing the Wigner functions in color space, we have
\begin{eqnarray}
\label{Jmu-Jn-2-I}
{\mathscr{J}}^{(2)I}_\mu &=& p_\mu   \frac{{\mathscr{J}}_n^{(2)I} }{p_n}
+\frac{s}{2p_n}\bar\epsilon_{\mu\alpha\beta}\partial_x^\alpha   {\mathscr{J}}^{(1)I,\beta}
+\frac{s}{4 N p_n} \bar\epsilon_{\mu\alpha\beta} { F}^{a,\alpha}_{\lambda} \partial^\lambda_p \mathscr{J}^{(0)a,\beta} ,\\
\label{Jmu-Jn-2-a}
{\mathscr{J}}^{(2)a}_\mu &=& p_\mu   \frac{{\mathscr{J}}_n^{(2)a} }{p_n}
+\frac{s}{2p_n}\bar\epsilon_{\mu\alpha\beta}\mathscr{D}^{\alpha,ac}   {\mathscr{J}}^{(1)c,\beta}
+\frac{s}{2  p_n} \bar\epsilon_{\mu\alpha\beta} { F}^{a,\alpha}_{\lambda} \partial^\lambda_p \mathscr{J}^{(0)I,\beta}\nonumber\\
& &+\frac{s}{4  p_n} \bar\epsilon_{\mu\alpha\beta} d^{cba}{ F}^{b,\alpha}_{\lambda} \partial^\lambda_p \mathscr{J}^{(0)c,\beta}
-\frac{1}{8p_n}f^{bca}\partial^\lambda_p \left( { F}_{n \lambda}^c\mathscr{J}^{(0)b}_{\mu}
-{ F}_{\mu \lambda}^c \mathscr{J}^{(0)b}_{n} \right)
\end{eqnarray}
which satisfy the covariant CKEs
\begin{eqnarray}
\label{CKE-8-I-2}
0 &=& \partial_x^\mu {{ \mathscr{J}}}_{\mu}^{(2)I}
+\frac{1}{2N}{ F}_{\mu \nu}^a \partial^\nu_p \mathscr{J}^{(1)a,\mu},\\
\label{CKE-8-a-2}
0 &=& \mathscr{D}^{ab,\mu} {{ \mathscr{J}}}_{\mu}^{(2)b}
+ { F}_{\mu \nu}^a  \partial^\nu_p  \mathscr{J}^{(1)I,\mu}
+\frac{1}{2}d^{bca} { F}_{\mu \nu}^c \partial^\nu_p \mathscr{J}^{(1)b,\mu}\nonumber\\
& &-\frac{1}{8} f^{bca} (\mathscr{D}_\lambda^{ce} { F}_{\mu \nu}^e) \partial^\nu_p \partial^\lambda_p
\mathscr{J}^{(0)b,\mu}
\end{eqnarray}
We note that both the constitutive equations (\ref{Jmu-Jn-2-I}) and (\ref{Jmu-Jn-2-a})  and the kinetic equations (\ref{CKE-8-I-2}) and (\ref{CKE-8-a-2}) for color singlet and multiplet of the  chiral Wigner functions
have been mixed with each other at second  order.

\section{Non-Abelian CKEs in seven-dimension phase space}
\label{sec:ckt-7}

The Wigner functions and equations based on quantum field theory exhibit explicit Lorentz covariance, in which we treat the momentum $p_n$ and $\bar p_\mu$ on the same footing. However,
such Lorentz covariance results in the singular Dirac delta functions as given in Eqs.(\ref{Jn-0-delta}), (\ref{Jsn-onshell-1}), and (\ref{Jsn-onshell-2}),
which is unsuitable for direct numerical calculations. In this section, we integrate over $p_n$ and derive the CKE in seven-dimensional phase space. It should be noted that if we integrate the covariant Wigner functions over
$p_n$ from the beginning, we obtain
\begin{eqnarray}
\label{wigner-hat}
 W(x_n,\bar x, \bar p)\equiv \int d p_n W(x,p)=\int\frac{d^3 \bar y}{(2\pi)^3} e^{-i\bar p\cdot \bar y}\left \langle  \rho \left(x+\frac{\bar y}{2},x-\frac{\bar y}{2}\right)\right \rangle,
\end{eqnarray}
which is simply the equal-time  (or single-time)  Wigner function at $x_n$. For this reason,  Wigner functions or their corresponding equations after integration over $p_n$
are often referred to as equal-time Wigner functions or equations.

\subsection{Zeroth-order result}

Instead of integrating over Eq.(\ref{Jmu-Jn-0}) directly, we integrate it with a arbitrary weighting function $z(p_n)$
because we always encounter such similar integration  with a specific weighting function. The integration is trivial at zeroth order
\begin{eqnarray}
\label{Jmu-0-int-g}
\int dp_n z(p_n) \mathscr{J}^{(0)}_\mu
 &=&\frac{1}{2} \sum_\lambda  z(\lambda|\bar p|)\dot x_\mu^{(0)}(\lambda) \mathcal{J}^{(0)}_{n}(\lambda)
\end{eqnarray}
where we have used
\begin{eqnarray}
\label{Dirac-delta}
\delta\left(p^2\right)&=&\frac{1}{2|\bar p|} \sum_\lambda \delta\left(p_n-\lambda|\bar p|\right).
\end{eqnarray}
and  $\mathcal{J}^{(0)}_{sn}(\lambda)$ and $\dot x_\mu^{(0)}(\lambda)$  are defined by
\begin{eqnarray}
 \mathcal{J}^{(0)}_{n}(\lambda ) \equiv  \left.\mathcal{J}^{(0)}_{n}\right|_{p_n=\lambda |\bar p|},\ \ \
 \dot{x}_\mu^{(0)}(\lambda) = \lambda n_\mu + \hat{\bar p}_\mu,\ \ \ \  \hat{\bar p}_\mu =\frac{ \bar p_\mu}{|\bar p|}
\end{eqnarray}
The summation index $\lambda$ has only two possibilities $+1$ and $-1$ where  $\lambda= +1$ denotes  positive energy contribution and $\lambda= -1$   negative energy contribution.
We can obtain the CKE in seven-dimensional phase space by integrating over Eq.(\ref{cke-0-8d}). The kinetic equation for the positive energy part can
be obtained by integrating from $0$ to $+\infty$, while the negative energy part from  $-\infty$ to $0$. Two branches can be written in a unified form
\begin{eqnarray}
\label{Jsn-eq-0-int}
0= \mathscr{D}_\mu\left( \dot{x}_\mu^{(0)}\mathcal{J}^{(0)}_{n} \right) &=&  \dot{x}_\mu^{(0)}\mathscr{D}_\mu\mathcal{J}^{(0)}_{n},
\end{eqnarray}
Decomposing the integral (\ref{Jmu-0-int-g}) in color space,  we obtain,
\begin{eqnarray}
\label{Jmu-I-0-int-z}
\int dp_n  z(p_n) \mathscr{J}_{s\mu}^{(0)I}
&=& \frac{1}{2} \sum_\lambda z \dot x_\mu^{(0)}
\mathcal{J}^{(0)I}_{sn} ,\\
\label{Jmu-a-0-int-z}
\int dp_n  z(p_n) \mathscr{J}_{s\mu}^{(0)a}
&=& \frac{1}{2} \sum_\lambda z \dot x_\mu^{(0)} \mathcal{J}^{(0)a}_{sn}
\end{eqnarray}
For brevity, we have omitted all the arguments associated with $\lambda$ in $ \mathcal{J}^{(0)}_{n}$, $\dot{x}^{(0)\mu}$,
and $z$ above and will also omit them in the following.  After integrating  the covariant CKEs  (\ref{cke-0-8d-I}) and (\ref{cke-0-8d-a}) over $p_n$ from $0$ to $+\infty$ or  from  $-\infty$ to $0$, we obtain the CKEs for
color singlet and mutiplet in seven-dimensional phase space
\begin{eqnarray}
\label{cke-0-7-f}
0 &=&\dot{x}^{(0)\mu}\partial_\mu^x \mathcal{J}^{(0)I}_{sn} ,\nonumber\\
0 &=&\dot{x}^{(0)\mu} {\mathscr{D}}^{ac}_\mu   \mathcal{J}^{(0)c}_{sn}
\end{eqnarray}

\subsection{First-order result}

Similarly,  we also integrate Eq.(\ref{Jmu-Jn-1}) with an arbitrary weighting function $z(p_n)$. The calculation is still trivial
\begin{eqnarray}
\label{Jmu-1-int-g}
\int dp_n z(p_n) \mathscr{J}^{(1)}_\mu
 &=&\frac{1}{2} \sum_\lambda  z\left(\dot x_\mu^{(0)} \mathcal{J}^{(1)}_{n} + \dot x_\mu^{(1)} \mathcal{J}^{(0)}_{n}\right)
\end{eqnarray}
where $ \dot x_\mu^{(1)}$ can be viewed as first-order correction to $ \dot x_\mu^{(0)}$
\begin{eqnarray}
\dot{x}_\mu^{(1)} = \frac{\lambda s}{2|\bar p|^2}\bar\epsilon_{\mu\alpha\beta}\bar p^\beta \mathscr{D}^\alpha.
\end{eqnarray}
We note that $ \dot x_\mu^{(1)}$  is  actually an operator instead of a usual function. It is also trivial to decomposing the integral (\ref{Jmu-1-int-g}) in color space:
\begin{eqnarray}
\label{Jmu-I-1-int-z}
\int dp_n z(p_n)  \mathscr{J}_{\mu}^{(1)I}
&=& \frac{1}{2} \sum_\lambda z
\left(\dot x_\mu^{(0)} \mathcal{J}^{(1)I}_{n} +  \dot{x}_\mu^{(1)I} \mathcal{J}^{(0)I}_{n} \right),\\
\label{Jmu-a-1-int-z}
\int dp_n z(p_n) \mathscr{J}_{\mu}^{(1)a}
&=& \frac{1}{2} \sum_\lambda z
\left(\dot x_\mu^{(0)} \mathcal{J}^{(1)a}_{n} +{\dot x}^{(1)ac}_\mu   \mathcal{J}^{(0)c}_{n}  \right),
\end{eqnarray}
where we have decomposed $ \dot x_\mu^{(1)}$ in color space
\begin{eqnarray}
\dot{x}_\mu^{(1)I} = \frac{\lambda s}{2|\bar p|^2}\bar\epsilon_{\mu\alpha\beta}\bar p^\beta \partial_x^\alpha,\ \ \ \
\dot{x}_\mu^{(1)ac} = \frac{\lambda s}{2|\bar p|^2}\bar\epsilon_{\mu\alpha\beta}\bar p^\beta \mathscr{D}^{ ac,\alpha} ,\ \ \
\end{eqnarray}
Let us integrate the covariant CKE  (\ref{cke-1-8d-I}) and (\ref{cke-1-8d-a}) over $p_n$,

\begin{eqnarray}
\label{cke-1-7d-I}
0 &=& \partial^x_\mu \left(\dot x^{(0)\mu} \mathcal{J}^{(1)I}_{n} +  \dot{x}^{(1)I,\mu} \mathcal{J}^{(0)I}_{n} \right)
 +\nabla_\mu^{(1)I a} \left( \dot x^{(0)\mu} \mathcal{J}^{(0)a}_{n}\right) \\
\label{cke-1-7d-a}
0 &=& {\mathscr{D}}^{ac}_\mu \left(\dot x^{(0)\mu} \mathcal{J}^{(1)c}_{n} +{\dot x}^{(1)cb,\mu }  \mathcal{J}^{(0)b}_{n}  \right)
 +  \nabla_\mu^{(1)a}\left( \dot x^{(0)\mu} \mathcal{J}^{(0)I}_{n}\right)
+\nabla_\mu^{(1) ac} \left(  \dot x^{(0)\mu} \mathcal{J}^{(0)c}_{n}\right)\hspace{1cm}
\end{eqnarray}
where we have defined some operators for brevity of symbol
\begin{eqnarray}
\nabla_\mu^{(1)I a} =
\frac{1}{2N}  {F}_{\mu \nu}^a \bar\partial^\nu_p, \ \ \
\nabla_\mu^{(1)a} =
 {F}_{\mu \nu}^a \bar\partial^\nu_p, \ \ \
\nabla_\mu^{(1)ac} =
\frac{1}{2}d^{cba}  {F}_{\mu \nu}^b \bar\partial^\nu_p, \ \ \
\end{eqnarray}
which can be transformed into the conventional  CKEs in seven-dimensional
phase space
\begin{eqnarray}
\label{cke-1-7-f}
0 &=& \dot{x}^{(0)\mu}\partial_\mu^x \mathcal{J}^{(1)I}_{sn}
+\frac{1}{2N} \dot{x}^{(0)\mu}  {F}_{\mu \nu}^a \bar\partial^\nu_p \mathcal{J}^{(0)a}_{sn},\nonumber\\
0 &=& \dot{x}^{(0)\mu}{\mathscr{D}}^{ac}_\mu   \mathcal{J}^{(1)c}_{sn}
+ \frac{1}{2}\dot{x}^{(0)\mu} d^{cba}  {F}_{\mu \nu}^b \bar\partial^\nu_p   \mathcal{J}^{(0)c}_{sn}
+ \frac{\lambda s}{4|\bar p|^2}f^{bca}\bar\epsilon^{\mu\alpha\beta}\bar p_\beta  F^b_{\mu\alpha}  \mathcal{J}^{(0)c}_{sn}\nonumber\\
& &+\dot{x}^{(0)\mu}{F}_{\mu \nu}^a \bar\partial^\nu_p   \mathcal{J}^{(0)I}_{sn}.
\end{eqnarray}

\subsection{Second-order result}

Now let us derive the second-order results in seven-dimensional phase space. Compared with the zeroth- or first-order result, the second-order result  would be  much more complicated
due to more complex expressions in Eq. (\ref{Jsn-onshell-2}), especially, in which  the derivatives of  Dirac delta function has been involved.
Let us first integrate  the Wigner function (\ref{Jsn-onshell-2}) with an arbitrary weighting function $z(p_n)$. The calculation is straightforward and the result reads
\begin{eqnarray}
\label{Jsn-2-int-g}
& &\int dp_n z(p_n) \mathscr{J}^{(2)}_{n}\nonumber\\
&=&\hspace{10pt}\frac{1}{2} \sum_\lambda  z \lambda\left[\mathcal{J}^{(2)}_{n}
+ \frac{s \bar\epsilon_{\mu\alpha\beta}\bar p^\beta}{8|\bar p|^2}
 \left\{ \mathcal{J}^{(0)\prime}_{n}, {{ F}^{\alpha\mu}} \right\}
 +\frac{\bar p^\mu \bar p^\nu - \bar p^2 \Delta^{\mu\nu}}{8|\bar p|^3}
\bar{\mathscr{D}}_\mu \bar{\mathscr{D}}_\nu \mathcal{J}^{(0)\prime}_{n}\right. \nonumber\\
& &\hspace{2.5cm}\left. -\frac{i \bar p^\mu}{8|\bar p|^2}
\left[\mathcal{J}_{s n}^{(0)\prime}, { F}_{\mu n}  \right]
+ \bar\partial^\nu_p   \left(\frac{i\lambda p^\mu(\lambda)}{8|\bar p|}
\left[\mathcal{J}_{n}^{(0)\prime}, { F}_{\mu \nu}  \right] \right)\right] \nonumber\\
& &-\frac{1}{2} \sum_\lambda  z \left[
\frac{ s\bar\epsilon_{\mu\alpha\beta}\bar p^\beta}{8|\bar p|^3}  \left\{ \mathcal{J}^{(0)}_{n}, {{ F}^{\alpha\mu}} \right\}
+\frac{\lambda \bar p^\mu \bar p^\nu - \bar p^2 \Delta^{\mu\nu}}{4|\bar p|^4}
 \bar{\mathscr{D}}_\mu \bar{\mathscr{D}}_\nu \mathcal{J}^{(0)}_{n} \right.\nonumber\\
& &\left.\hspace{2.5cm}-\frac{i \bar p^\mu }{8|\bar p|^3}\left[\mathcal{J}_{n}^{(0)}, { F}_{\mu n}  \right]
-\bar\partial^\nu_p \left(\frac{i}{8|\bar p|}\left[\mathcal{J}_{n}^{(0)}, { F}_{n \nu}  \right] \right)\right]\nonumber\\
& &+\frac{1}{2} \sum_\lambda  z' \lambda \left[
\frac{ s\bar\epsilon_{\mu\alpha\beta}\bar p^\beta }{8|\bar p|^2} \left\{ \mathcal{J}^{(0)}_{n}, {{ F}^{\alpha\mu}} \right\}
+\frac{\lambda \bar p^\mu \bar p^\nu - \bar p^2\Delta^{\mu\nu}}{8|\bar p|^3}
\bar{\mathscr{D}}_\mu \bar{\mathscr{D}}_\nu \mathcal{J}^{(0)}_{n} \right.\nonumber\\
& &\left.\hspace{2.5cm}+\bar\partial^\nu_p \left(\frac{i (|\bar p|n^\mu + \lambda\bar p^\mu)}{8|\bar p|}
\left[\mathcal{J}_{n}^{(0)}, { F}_{\mu \nu}  \right] \right)\right]
\end{eqnarray}
where the  functions with primes denotes derivative and are defined by
\begin{eqnarray}
z'\equiv z'(\lambda|\bar p|)=\left. \frac{\partial z(p_n)}{\partial p_n}\right|_{p_n =\lambda |\bar p|},\ \ \
 \mathcal{J}_{s n}^{(0)\prime} = \left. \frac{\partial  \mathcal{J}_{s n}^{(0)} }{\partial p_n}\right|_{p_n =\lambda |\bar p|}.
\end{eqnarray}
To arrive at the result (\ref{Jsn-2-int-g}), we have used the identity  (\ref{Dirac-delta}) and
\begin{eqnarray}
\delta'\left(p^2\right)&=&\frac{1}{4|\bar p|^2}\sum_\lambda \left[\lambda \delta'\left(p_n-\lambda|\bar p|\right)
+\frac{1}{|\bar p|}\delta\left(p_n-\lambda|\bar p|\right)\right]
\end{eqnarray}
together with some right integration by parts.
After integration, $\mathcal{J}_{ n}^{(0)\prime}$ will not depend on $\mathcal{J}_{ n}^{(0)}$  and we can not obtain $\mathcal{J}_{ n}^{(0)\prime}$
 from $\mathcal{J}_{ n}^{(0)}$. Fortunately, $\mathcal{J}_{ n}^{(0)\prime}$ always appears along with the weighting function $z$ instead of
$ z'$. Therefore, we can lump all the terms associated with $\mathcal{J}_{n}^{(0)\prime}$ into the distribution $ \mathcal{J}^{(2)}_{n}$ by redefinition:
\begin{eqnarray}
\label{redefinition}
\mathcal{J}^{(2)}_{n} & \leftarrow & \mathcal{J}^{(2)}_{n}
+ \frac{s}{8|\bar p|^2} \bar\epsilon_{\mu\alpha\beta}\bar p^\beta
 \left\{ \mathcal{J}^{(0)\prime}_{n}, {{ F}^{\alpha\mu}} \right\}
+\frac{1}{8|\bar p|^3}\left(\bar p^\mu \bar p^\nu - \bar p^2 \Delta^{\mu\nu}\right)
\bar{\mathscr{D}}_\mu \bar{\mathscr{D}}_\nu \mathcal{J}^{(0)\prime}_{n}\nonumber\\
& & -\frac{i}{8|\bar p|^2} \bar p^\mu
\left[\mathcal{J}_{n}^{(0)\prime}, { F}_{\mu n}  \right]
+ \bar\partial^\nu_p   \left(\frac{i\lambda}{8|\bar p|} p^\mu(\lambda)
\left[\mathcal{J}_{n}^{(0)\prime}, { F}_{\mu \nu}  \right] \right)
\end{eqnarray}
For brevity, we still use the same function $\mathcal{J}^{(2)}_{n} $ to represent the redefined one.
With this redefinition, we can integrate the Eq.(\ref{Jmu-Jn-2}) over $p_n$,
\begin{eqnarray}
\label{Jmu-Jn-2-int}
\int dp_n {\mathscr{J}}^{(2)}_\mu &=& n_\mu \int dp_n {\mathscr{J}}^{(2)}_{n} +\bar p_\mu  \int dp_n \frac{1 }{p_n} {\mathscr{J}}^{(2)}_{n}
+\frac{s}{2}\bar\epsilon_{\mu\alpha\beta} \mathscr{D}^\alpha\int dp_n \frac{1}{p_n} {\mathscr{J}}^{(1)\beta}\nonumber\\
& &+\frac{s}{4}\bar\epsilon_{\mu\alpha\beta} \bar\partial^\lambda_p \left\{\int dp_n \frac{1}{p_n}\mathscr{J}^{(0)\beta}, {{ F}^\alpha}_{\lambda} \right\}
+\frac{s}{4}\bar\epsilon_{\mu\alpha\beta} \left\{\int dp_n \frac{1}{p_n^2}\mathscr{J}^{(0)\beta}, {{ F}^\alpha}_{n} \right\}\nonumber\\
& & +\frac{i}{8}\bar\partial^\lambda_p \left(\left[\int dp_n \frac{1}{p_n}\mathscr{J}^{(0)}_{\mu}, { F}_{n \lambda}\right]
- \left[\int dp_n \frac{1}{p_n} \mathscr{J}^{(0)}_{n}, { F}_{\mu \lambda}\right]  \right)\nonumber\\
& &-\frac{i}{8} \left[\int dp_n \frac{1}{p_n^2}\mathscr{J}^{(0)}_{n}, { F}_{\mu \lambda}\right]
\end{eqnarray}
The integrals in the first and second terms on the right-handed side of the equation above can be finished by using the general result (\ref{Jsn-2-int-g}) with specific assignments $z(p_n)=1$ and $z(p_n)=1/p_n$, respectively.
The integrals in the third term can be calculated with (\ref{Jmu-1-int-g}) by setting $z(p_n)=1/p_n$. We can integrate the fourth and sixth terms by using the result (\ref{Jmu-0-int-g})  with $z(p_n)=1/p_n$ and
the fifth and final terms with $z(p_n)=1/p_n^2$. The final result reads

\begin{eqnarray}
\int dp_n\mathscr{J}_{\mu}^{(2)}
&=&\hspace{10pt} \frac{1}{2} \sum_\lambda  {\dot x}^{(0)}_\mu \mathcal{J}^{(2)}_{n}
 +\frac{1}{2} \sum_\lambda \lambda \frac{ s  \bar\epsilon_{\mu\alpha\beta}\bar p^\beta}{2|\bar p|^2}
 \mathscr{D}^\alpha \mathcal{J}^{(1)}_{n} \nonumber\\
& &-\frac{1}{2} \sum_\lambda \lambda  \left[
\frac{ s \bar\epsilon_{\gamma\alpha\beta}\bar p^\beta}{8|\bar p|^3}
\left\{\hat{\bar p}_\mu \mathcal{J}^{(0)}_{n}, {{ F}^{\alpha\gamma}} \right\}
+\frac{\lambda \left(\bar p^\alpha \bar p^\beta - \bar p^2 \Delta^{\alpha\beta}\right)}{4|\bar p|^4}
 \bar{\mathscr{D}}_\alpha \bar{\mathscr{D}}_\beta \hat{\bar p}_\mu \mathcal{J}^{(0)}_{n} \right.\nonumber\\
& &\left.\hspace{2.0cm}-\frac{i}{8|\bar p|^3} \bar p^\alpha \left[\hat{\bar p}_\mu \mathcal{J}_{n}^{(0)}, { F}_{\alpha n}  \right]
-\bar\partial^\alpha_p \left(\frac{i}{8|\bar p|}
\left[\hat{\bar p}_\mu\mathcal{J}_{n}^{(0)}, { F}_{n \alpha}  \right] \right)\right]\nonumber\\
& &-\frac{1}{2} \sum_\lambda  \lambda  \left[
\frac{ s \bar\epsilon_{\gamma\alpha\beta}\bar p^\beta}{8|\bar p|^3}
\left\{\hat{\bar p}_\mu  \mathcal{J}^{(0)}_{n}, {{ F}^{\alpha\gamma}} \right\}
- \frac{s\lambda\bar\epsilon_{\mu\alpha\beta}\bar p^\beta }{4|\bar p|^3}
\left\{ \mathcal{J}^{(0)}_{n},  {{ F}^{\alpha n}} \right\}
\right]\nonumber\\
& & - \frac{1}{2} \sum_\lambda \frac{ 1 }{8|\bar p|^5}
\left[ \bar p_\alpha \bar p_\beta\bar p_\mu
+ \bar p^2\left( \Delta_{\alpha\beta}\bar p_\mu
-2\Delta_{\mu\beta}\bar p_\alpha \right) \right]
 \mathscr{D}^\alpha  \mathscr{D}^\beta \mathcal{J}^{(0)}_{n} \nonumber\\
& &-\frac{1}{2} \sum_\lambda  \lambda
 \frac{i \left(\hat{\bar p}^\alpha\hat{\bar p}_\mu n^\nu -\Delta^\nu_\mu n^\alpha -\lambda \hat{\bar p}^\alpha \Delta^\nu_\mu\right)}{8|\bar p|^2}
\left[ \mathcal{J}_{n}^{(0)}, { F}_{\alpha \nu}  \right]
\nonumber\\
& &  +\frac{1}{2} \sum_\lambda \lambda \bar\epsilon_{\mu\alpha\beta} \bar\partial^\rho_p
\left(\frac{s \bar p^\beta}{4|\bar p|^2}\left\{ \mathcal{J}^{(0)}_{n}, {{ F}^\alpha}_{\rho} \right\}\right)
\nonumber\\
& &-\frac{1}{2} \sum_\lambda \bar\partial^\beta_p\left[
\frac{i}{8|\bar p|^3}\left({\bar p}_\mu \bar p^\alpha -\bar p^2 \Delta^\alpha_\mu\right)
\left[\mathcal{J}_{ n}^{(0)}, { F}_{\alpha \beta}  \right] \right]
\end{eqnarray}
Note that the second-order distribution function $\mathcal{J}^{(2)}_{n}$ above has already been  the newly defined  distribution function via (\ref{redefinition}).
Decomposing this expression in color space, we can rewrite the results in  more systematic forms,
\begin{eqnarray}
\label{Jmu-I-int}
\int dp_n  \mathscr{J}_{\mu}^{(2)I}
&=& \frac{1}{2} \sum_\lambda \left(
\dot x_\mu^{(0)} \mathcal{J}^{(2)I}_{n} +  \dot{x}_\mu^{(1)I} \mathcal{J}^{(1)I}_{n}
 + \sqrt{\omega}^{(2)I} \dot{ x}_\mu^{(0)} \mathcal{J}^{(0)I}_{n}
+ \sqrt{\omega}^{(2)Ia} \dot{ x}_\mu^{(0)} \mathcal{J}^{(0)a}_{n} \right.\nonumber\\
& &\left.\hspace{1.2cm} + \dot{ x}_\mu^{(2)I} \mathcal{J}^{(0)I}_{n}
+  \dot{ x}_\mu^{(2)I a} \mathcal{J}^{(0)a}_{n} \right),\\
\label{Jmu-a-int}
\int dp_n  \mathscr{J}_{\mu}^{(2)a}
&=& \frac{1}{2} \sum_\lambda \left(
\dot x_\mu^{(0)} \mathcal{J}^{(2)a}_{n} +{\dot x}^{(1)ab}_\mu   \mathcal{J}^{(1)b}_{n}
 + \sqrt{\omega}^{(2)a }\dot{ x}_\mu^{(0)}  \mathcal{J}^{(0)I}_{n}
+ \sqrt{\omega}^{(2)ab} \dot{ x}_\mu^{(0)} \mathcal{J}^{(0)b}_{n}  \right.\nonumber\\
& &\left.\hspace{1.2cm} + \dot{ x}_\mu^{(2)a}  \mathcal{J}^{(0)I}_{n}
+\dot{ x}_\mu^{(2)ab} \mathcal{J}^{(0)b}_{n} \right)
\end{eqnarray}
where  $\dot{{x}}^{(2)\mu}$ denotes second-order correction to the four-velocity $\dot{x}^{(0)\mu}$
\begin{eqnarray}
\dot{\bar x}_\mu^{(2)I}&=&
 - \frac{ 1 }{8|\bar p|^5}
\left[ \bar p_\alpha \bar p_\beta\bar p_\mu
+ \bar p^2\left( \Delta_{\alpha\beta}\bar p_\mu
-2\Delta_{\mu\beta}\bar p_\alpha \right) \right]
 \partial_x^\alpha  \partial_x^\beta ,\\
\dot{\bar x}_\mu^{(2)Ia}&=& \frac{\dot{\bar x}_\mu^{(2)a}}{2N}
= - \frac{ \lambda s \hat{\bar p}_\mu \bar\epsilon_{\gamma\alpha\beta}\bar p^\beta}{8N|\bar p|^3}
{{ F}^{a,\alpha\gamma}}
+ \frac{s\bar\epsilon_{\mu\alpha\beta}\bar p^\beta }{4N|\bar p|^3}  {{ F}^{a,\alpha n}}
 +  \bar\partial_\beta^p
\left(\frac{\lambda  s \bar\epsilon_{\mu\alpha\nu}\bar p^\nu }{4N|\bar p|^2} { F}^{a,\alpha\beta} \right),  \\
\dot{\bar x}_\mu^{(2)ab}&=&
- \frac{ \lambda s \bar\epsilon_{\gamma\alpha\beta}\bar p^\beta}{8|\bar p|^3}
d^{bca}{{ F}^{c,\alpha\gamma}} \hat{\bar p}_\mu
+ \frac{s\bar\epsilon_{\mu\alpha\beta}\bar p^\beta }{4|\bar p|^3} d^{bca} {{ F}^{c,\alpha n}}\nonumber\\
& &+  \frac{\lambda }{8|\bar p|^3}
\left(\hat{\bar p}^\alpha\hat{\bar p}_\mu n^\nu -\Delta^\nu_\mu n^\alpha -\lambda \hat{\bar p}^\alpha\Delta^\nu_\mu\right)
f^{bca}{ F}_{\alpha \nu}^c
\nonumber\\
& & - \frac{ 1 }{8|\bar p|^5}
\left[ \bar p_\alpha \bar p_\beta\bar p_\mu
+ \bar p^2\left( \Delta_{\alpha\beta}\bar p_\mu
-2\Delta_{\mu\beta}\bar p_\alpha \right) \right]
{\mathscr{D}}^{ac,\alpha} {\mathscr{D}}^{cb,\beta}  \nonumber\\
& &  +  \bar\partial_\beta^p
\left(\frac{\lambda  s \bar\epsilon_{\mu\alpha\nu}\bar p^\nu }{4|\bar p|^2} d^{bca} { F}^{c,\alpha\beta}
+\frac{{\bar p}_\mu \bar p_\alpha -\bar p^2 \Delta_{\alpha\mu}}{8|\bar p|^3}f^{bca} {F}^{c,\alpha \beta} \right)
\end{eqnarray}
We should view these expressions  as operators and the derivatives in the last terms of $\dot{\bar x}_\mu^{(2)Ia}$ and $\dot{\bar x}_\mu^{(2)ab}$
act on not only the terms in the brackets but also the distribution function  $\mathcal{J}^{(0)b}_{n} $.
The factor $\sqrt{\omega}^{(2)}$ can be viewed as the second-order correction to the invariant phase space
\begin{eqnarray}
\sqrt{\omega}^{(2) I}&=& -\frac{1}{4|\bar p|^4}\left(\bar p^\alpha \bar p^\beta -\bar p^2 \Delta^{\alpha\beta}\right)\partial^x_\alpha \partial^x_\beta,\\
\sqrt{\omega}^{(2)I a }&=& \frac{\sqrt{\omega}^{(2) a }}{2N} =  -\frac{\lambda  s}{8 N |\bar p|^3}\bar \epsilon^{\nu\alpha\beta}\bar p_\beta F_{\alpha\nu}^a,\\
\sqrt{\omega}^{(2) ab }&=&
-\frac{\lambda s}{8|\bar p|^3} \bar\epsilon_{\nu\alpha\beta}\bar p^\beta    d^{bca}F^{c,\alpha\nu}
-\frac{1}{4|\bar p|^4}\left(\bar p^\alpha \bar p^\beta - \bar p^2 \Delta^{\alpha\beta}\right)
\bar{\mathscr{D}}^{ac}_\alpha \bar{\mathscr{D}}^{cb}_\beta  \nonumber\\
& &-\frac{\lambda }{8|\bar p|^3}f^{bca}  F_{\mu n}^c  \bar p^\mu
-\bar\partial^\nu_p \left(\frac{\lambda }{8|\bar p|}f^{bca}  F_{n \nu}^c \right),
\end{eqnarray}
which  can be seen more clearly if we only keep the time-like  component in (\ref{Jmu-I-int}) and (\ref{Jmu-a-int})
\begin{eqnarray}
\int dp_n  \mathscr{J}_{n}^{(2)I}
&=& \frac{1}{2} \sum_\lambda  \lambda \left(
 \mathcal{J}^{(2)I}_{n}
 + \sqrt{\omega}^{(2)I} \mathcal{J}^{(0)I}_{n}
+ \sqrt{\omega}^{(2)Ia} \mathcal{J}^{(0)a}_{n}  \right),\\
\int dp_n  \mathscr{J}_{n}^{(2)a}
&=& \frac{1}{2} \sum_\lambda \lambda \left(
 \mathcal{J}^{(2)a}_{n}
 + \sqrt{\omega}^{(2)a } \mathcal{J}^{(0)I}_{n}
+ \sqrt{\omega}^{(2)ab} \mathcal{J}^{(0)b}_{n} \right)
\end{eqnarray}
We did not show  the lower-order factor for the invariant phase space explicitly
because they are both trivial with  $\sqrt{\omega}^{(1)} = 0$ and $\sqrt{\omega}^{(0)}=1$.  Note that the factors associated with $\sqrt{\omega}^{(2)}$ are in general all operators.
Similarly, the derivative in the last term of $\sqrt{\omega}^{(2) ab }$ acts on all the terms after it including the distribution function $\mathcal{J}^{(0)b}_{n}$.

The CKE in seven-dimensional phase space can be derived by integrating Eqs.(\ref{CKE-8-I-2}) and (\ref{CKE-8-a-2}) over $p_n$. Using the results
(\ref{Jmu-I-int}), (\ref{Jmu-a-int}), (\ref{Jmu-I-1-int-z}), (\ref{Jmu-a-1-int-z}), and (\ref{Jmu-a-0-int-z}), we have
\begin{eqnarray}
\label{cke-2-7d-I}
0 &=& \partial^x_\mu \left(
\dot x_\mu^{(0)} \mathcal{J}^{(2)I}_{n} +  \dot{x}_\mu^{(1)I} \mathcal{J}^{(1)I}_{n}
 + \sqrt{\omega}^{(2)I} \dot{ x}_\mu^{(0)} \mathcal{J}^{(0)I}_{n}
+ \sqrt{\omega}^{(2)Ia} \dot{ x}_\mu^{(0)} \mathcal{J}^{(0)a}_{n} \right.\nonumber\\
& &\left. + \dot{ x}_\mu^{(2)I} \mathcal{J}^{(0)I}_{n}
+  \dot{ x}_\mu^{(2)I a} \mathcal{J}^{(0)a}_{n} \right)
 +\nabla_\mu^{(1)I a} \left(\dot x_\mu^{(0)} \mathcal{J}^{(1)a}_{n} +{\dot x}^{(1)ac}_\mu   \mathcal{J}^{(0)c}_{n}  \right) \\
\label{cke-2-7d-a}
0 &=& \mathscr{D}^{ab,\mu}\left(
\dot x_\mu^{(0)} \mathcal{J}^{(2)b}_{n} +{\dot x}^{(1)bc}_\mu   \mathcal{J}^{(1)c}_{n}
 + \sqrt{\omega}^{(2)b }\dot{ x}_\mu^{(0)}  \mathcal{J}^{(0)I}_{n}
+ \sqrt{\omega}^{(2)bc} \dot{ x}_\mu^{(0)} \mathcal{J}^{(0)c}_{n}  \right.\nonumber\\
& &\left. + \dot{ x}_\mu^{(2)b}  \mathcal{J}^{(0)I}_{n}
+\dot{ x}_\mu^{(2)bc} \mathcal{J}^{(0)c}_{n} \right)
+ \nabla_\mu^{(1)a} \left(\dot x_\mu^{(0)} \mathcal{J}^{(1)I}_{n} +  \dot{x}_\mu^{(1)I} \mathcal{J}^{(0)I}_{n} \right)
\nonumber\\
& &+\nabla_\mu^{(1) ab} \left(\dot x_\mu^{(0)} \mathcal{J}^{(1)b}_{n} +{\dot x}^{(1)bd}_\mu   \mathcal{J}^{(0)d}_{n}  \right)
\nabla_\mu^{(2)ab} \left(\dot x_\mu^{(0)} \mathcal{J}^{(1)b}_{n}  \right)
\end{eqnarray}
where we have defined the second-order operator
\begin{eqnarray}
\nabla_\mu^{(2)ac} &=&
-\frac{1}{8} f^{cba} \bar\partial^\nu_p \bar\partial^\lambda_p  \left(\mathscr{D}_\lambda^{be}{ F}^e_{\mu \nu}\right).
\end{eqnarray}
Following the approach  to put the chiral kinetic equations in conventional form as given in \cite{Yang:2024hmd}, we can move the operator $\nabla_\mu$ close to
the distribution function
\begin{eqnarray}
\label{cke-2-I-formal}
0&=& \dot{x}^{(0)\mu}\partial_\mu^x \mathcal{J}^{(2)I}_{n}
+\dot{x}^{(1)I,\mu} \partial_\mu^x \mathcal{J}^{(1)I}_{n}
+\dot{x}^{(2)I,\mu} \partial_\mu^x \mathcal{J}^{(0)I}_{n}
\nonumber\\
& & +\dot{x}^{(0)\mu}\nabla_\mu^{(1)Ic} \mathcal{J}^{(1)c}_{n}
+\dot{x}^{(1)I,\mu}\nabla_\mu^{(1)Ic} \mathcal{J}^{(0)c}_{n}
+\dot{x}^{(2)Ia,\mu}{\mathscr{D}}^{ac}_\mu   \mathcal{J}^{(0)c}_{n}
\nonumber\\
& &+\left(\nabla^{(1)Ib}_\mu{\dot x}^{(1)bc,\mu} -\dot{x}^{(1)I,\mu}\nabla_\mu^{(1)Ic}  \right) \mathcal{J}^{(0)c}_{n}\nonumber\\
& &+\left({\mathscr{D}}^{ca}_\mu \dot{x}^{(2)Ia,\mu}\right)   \mathcal{J}^{(0)c}_{n}
+\left({\mathscr{D}}^{ca}_\mu\sqrt{\omega}^{(2)Ia} \right) \dot{x}^{(0)\mu}   \mathcal{J}^{(0)c}_{n},\\
\label{cke-2-a-formal}
0&=&\dot{x}^{(0)\mu}{\mathscr{D}}^{ac}_\mu   \mathcal{J}^{(2)c}_{n}
+\dot{x}^{(1)ab,\mu}{\mathscr{D}}^{bc}_\mu   \mathcal{J}^{(1)c}_{n}
+ \dot{x}^{(2)ab,\mu}{\mathscr{D}}^{bc}_\mu   \mathcal{J}^{(0)c}_{n}
\nonumber\\
& &+\dot{x}^{(0)\mu}{\nabla}^{(1)ac}_\mu   \mathcal{J}^{(1)c}_{n}
 +\dot{x}^{(1)ab,\mu}{\nabla}^{(1)bc}_\mu   \mathcal{J}^{(0)c}_{n}
 + \dot{x}^{(0)\mu} \nabla_\mu^{(2)ac}\mathcal{J}^{(0)c}_{n}
\nonumber\\
& &+ \left(\mathscr{D}^{ab}_\mu   {\dot x}^{(1)bc,\mu}
-\dot{x}^{(1)ab,\mu}{\mathscr{D}}^{bc}_\mu \right)  \mathcal{J}^{(1)c}_{n}
+\left( \nabla^{(1)a b}_\mu {\dot x}^{(1)bc,\mu}  -\dot{x}^{(1)ab,\mu}{\nabla}^{(1)bc}_\mu   \right)\mathcal{J}^{(0)c}_{n} \nonumber\\
& & +\left(\nabla_\mu^{(2)ac} \dot{x}^{(0)\mu}-\dot{x}^{(0)\mu} \nabla_\mu^{(2)ac}\right) \mathcal{J}^{(0)c}_{n}
+\left(\mathscr{D}^{ab,\mu}\dot{ x}_\mu^{(2)bc}
- \dot{x}^{(2)ab,\mu}{\mathscr{D}}^{bc}_\mu    \right)\mathcal{J}^{(0)c}_{n}\nonumber\\
& &+\left(\mathscr{D}^{ab,\mu} \sqrt{\omega}^{(2)bc} \dot{ x}_\mu^{(0)}-\sqrt{\omega}^{(2)ab}\dot{x}^{(0)\mu}{\mathscr{D}}^{bc}_\mu   \right)
 \mathcal{J}^{(0)c}_{n}\nonumber\\
& &+\dot{x}^{(0)\mu}{\nabla}^{(1) a}_\mu   \mathcal{J}^{(1)I}_{n}
+\dot{x}^{(1)ab,\mu} {\nabla}^{(1) b}_\mu   \mathcal{J}^{(0)I}_{n}
+ \dot{x}^{(2)a,\mu} \partial_\mu^x \mathcal{J}^{(0)I}_{n}
\nonumber\\
& &+ \left( \nabla^{(1)a}_\mu \dot{x}^{(1)I,\mu} -\dot{x}^{(1)ab,\mu}{\nabla}^{(1) b}_\mu     \right) \mathcal{J}^{(0)I}_{n}\nonumber\\
& &+\left(\mathscr{D}^{ab,\mu}\dot{ x}_\mu^{(2)b} \right) \mathcal{J}^{(0)I}_{n}
 +\left(\mathscr{D}^{ab,\mu} \sqrt{\omega}^{(2)b }\dot{ x}_\mu^{(0)}  \right)\mathcal{J}^{(0)I}_{n}
\end{eqnarray}
where  we have used the zeroth- and first-order kinetic equations (\ref{cke-0-7-f}) and (\ref{cke-1-7-f}). After direct but lengthy calculation, we obtain the final results:
\begin{eqnarray}
\label{cke-I-2-final}
0&=& \dot{x}^{(0)\mu}\partial_\mu^x \mathcal{J}^{(2)I}_{sn}
+\frac{1}{2N}  \dot{x}^{(0)\mu}  {F}_{\mu \nu}^a \bar\partial^\nu_p \mathcal{J}^{(1)a}_{sn}\nonumber\\
& & - \frac{ 1 }{8|\bar p|^5} \left( \bar p^\alpha \bar p^\beta
- \bar p^2\Delta^{\alpha\beta} \right)\bar p^\mu
 \partial^x_\alpha  \partial^x_\beta \partial_\mu^x \mathcal{J}^{(0)I}_{sn}
\nonumber\\
& & -\frac{\lambda s}{4N|\bar p|^2}\bar\epsilon^{\mu\alpha\beta}\bar p_\beta\left( \mathscr{D}^{ab}_\mu
 {F}_{\nu\alpha}^b\right)  \bar\partial^\nu_p \mathcal{J}^{(0)a}_{sn}
\nonumber\\
& &
- \frac{  s}{8N|\bar p|^3}\left(\lambda \bar\epsilon^{\mu\alpha\beta} \hat{\bar p}^\nu -2\bar\epsilon^{\nu\alpha\beta}  n^\mu  \right)\bar p_\beta
{{ F}^{b}_{\alpha\mu}} {\mathscr{D}}^{ba}_\nu   \mathcal{J}^{(0)a}_{sn}
\end{eqnarray}
\begin{eqnarray}
\label{cke-a-2-final}
0&=&\dot{x}^{(0)\mu}{\mathscr{D}}^{ac}_\mu   \mathcal{J}^{(2)c}_{sn}
+\dot{x}^{(0)\mu} {F}_{\mu \nu}^a \bar\partial^\nu_p  \mathcal{J}^{(1)I}_{sn}
+\frac{1}{2} \dot{x}^{(0)\mu}d^{cba}  {F}_{\mu \nu}^b \bar\partial^\nu_p   \mathcal{J}^{(1)c}_{sn}
+ \frac{\lambda s}{4|\bar p|^2} \bar\epsilon^{\mu\alpha\beta}\bar p_\beta f^{bca} F^b_{\mu\alpha} \mathcal{J}^{(1)c}_{sn}
\nonumber\\
& & - \frac{ 1 }{24|\bar p|^5}\left[3 \bar p^\alpha \bar p^\beta \bar p^\mu -\bar p^2
\left(  \Delta^{\alpha\beta} \bar p^\mu + \Delta^{\beta\mu} \bar p^\alpha + \Delta^{\mu\alpha} \bar p^\beta  \right) \right]
{\mathscr{D}}^{ae}_\alpha {\mathscr{D}}^{eb}_\beta {\mathscr{D}}^{bc}_\mu   \mathcal{J}^{(0)c}_{sn}  \nonumber\\
& &+ \frac{  s}{8|\bar p|^3}\left(\lambda \bar\epsilon^{\mu\alpha\beta}\hat{\bar p}^\nu -2\bar\epsilon^{\nu\alpha\beta} n^\mu\right)  \bar p_\beta
d^{bea}{{ F}^{e}_{\mu\alpha}} {\mathscr{D}}^{bc}_\nu   \mathcal{J}^{(0)c}_{sn}
-\frac{\lambda s}{4|\bar p|^2}\bar\epsilon^{\mu\alpha\beta}\bar p_\beta
d^{cba}\left( \mathscr{D}^{be}_\alpha {F}^{e}_{\mu \nu} \right) \bar\partial^\nu_p   \mathcal{J}^{(0)c}_{sn}
\nonumber\\
& & -  \frac{\lambda}{8|\bar p|^2} \left(3\hat{\bar p}^\alpha \hat{\bar p}^\mu n^\nu -3n^{\alpha} \Delta^{\mu\nu} - 2\lambda \hat{\bar p}^\alpha \Delta^{\mu\nu}\right)
f^{bea}{ F}_{\alpha \nu}^e {\mathscr{D}}^{bc}_\mu   \mathcal{J}^{(0)c}_{sn}
\nonumber\\
& &+ \frac{ 1 }{6|\bar p|^3} \Delta^{\alpha\beta} \bar p^\mu
 f^{cba}\left( {\mathscr{D}}^{be}_{\alpha} F^e_{\mu\beta} \right) \mathcal{J}^{(0)c}_{sn}
 +  \frac{1}{8|\bar p|^3}\left({\bar p}^\mu \bar p^\alpha -\bar p^2 \Delta^{\alpha\mu}\right)
f^{bea} {F}^{e}_{\alpha \beta} {\mathscr{D}}^{bc}_\mu  \bar\partial^\beta_p \mathcal{J}^{(0)c}_{sn}
\nonumber\\
& & -\frac{1}{8} \dot{x}^{(0)\mu} f^{cba}   \left(\mathscr{D}_\lambda^{be}{ F}^e_{\mu \nu}\right)\bar\partial^\nu_p \bar\partial^\lambda_p\mathcal{J}^{(0)c}_{sn}
 -\frac{\lambda }{8|\bar p|}f^{cba}\dot{x}^{(0)\mu} \left({\mathscr{D}}^{be}_\mu  F_{n \nu}^e \right) \bar\partial^\nu_p  \mathcal{J}^{(0)c}_{sn}\nonumber\\
& &-\frac{\lambda s}{2|\bar p|^2}\bar\epsilon^{\mu\alpha\beta}\bar p_\beta\left( \mathscr{D}^{ ab}_\mu  {F}_{\nu \alpha}^b \right) \bar\partial^\nu_p  \mathcal{J}^{(0)I}_{sn}
 + \frac{s}{4|\bar p|^3}\left(\lambda  \bar\epsilon^{\nu\alpha\beta}\hat{\bar p}^\mu -2\bar\epsilon^{\mu\alpha\beta} n^\nu \right) \bar p_\beta
{{ F}^{a}_{\nu\alpha}}  \partial_\mu^x \mathcal{J}^{(0)I}_{sn}
\end{eqnarray}
We have present some calculation details in the appendix.

\subsection{Summing the results}
In  this section, we  recombine  the  perturbative results obtained above  into a unified form.
The  full distribution function up to second order is given  by
\begin{eqnarray}
\mathscr{J}^{\mu} &=& \mathscr{J}^{(0)\mu} +  \mathscr{J}^{(1)\mu} +  \mathscr{J}^{(2)\mu}
\end{eqnarray}
After  the integration of  the full $\mathscr{J}^{\mu}$ and decomposition in color space, we have
\begin{eqnarray}
\label{Jmu-I-int}
\int dp_n  \mathscr{J}_{\mu}^{I}
&=& \frac{1}{2} \sum_\lambda \left(
\sqrt{\omega}^{I} \dot x_\mu^I\mathcal{J}^{I}_{n}
+ \sqrt{\omega}^{Ia} \dot{ x}_\mu^I \mathcal{J}^{a}_{n}
+ \sqrt{\omega}^{I} \dot{ x}_\mu^{I a} \mathcal{J}^a_{n} \right),\\
\label{Jmu-a-int}
\int dp_n  \mathscr{J}_{\mu}^{a}
&=& \frac{1}{2} \sum_\lambda \left(\sqrt{\omega}^{ab} \dot{ x}_\mu^I \mathcal{J}^{b}_{n}
+ \sqrt{\omega}^{I} \dot{ x}_\mu^{ab} \mathcal{J}^{b}_{n}
 + \sqrt{\omega}^{a }\dot{ x}_\mu^I \mathcal{J}^{I}_{n}
 +\sqrt{\omega}^{I} \dot{ x}_\mu^{a}  \mathcal{J}^{I}_{n}
 \right)
\end{eqnarray}
where
\begin{eqnarray}
\mathcal{J}^{I}_{n}&=& \mathcal{J}^{(0)I}_{n}+\mathcal{J}^{(1)I}_{n}+\mathcal{J}^{(2)I}_{n},\ \ \
\mathcal{J}^{a}_{n}= \mathcal{J}^{(0)a}_{n}+\mathcal{J}^{(1)a}_{n}+\mathcal{J}^{(2)a}_{n},\\
\sqrt{\omega}^{I} &=& 1 +\sqrt{\omega}^{(2)I},\ \  \sqrt{\omega}^{Ia} =\sqrt{\omega}^{(2)Ia},\ \
\sqrt{\omega}^{a} =\sqrt{\omega}^{(2)a},\ \   \sqrt{\omega}^{ab} =\delta^{ab} +\sqrt{\omega}^{(2)ab}  ,\\
\dot{ x}_\mu^{I} &=&\dot{ x}_\mu^{(0)}+\dot{ x}_\mu^{(1)I} + \dot{ x}_\mu^{(2)I} ,\ \
\dot{ x}_\mu^{Ia} = \dot{ x}_\mu^{(2)Ia} ,\ \ \dot{ x}_\mu^{a} = \dot{ x}_\mu^{(2)a},\ \
\dot{ x}_\mu^{ab}= \dot{ x}_\mu^{(1)ab} + \dot{ x}_\mu^{(2)ab},
\end{eqnarray}
Note that the distribution functions  $\mathcal{J}^{(2)I}_{n}$ and $\mathcal{J}^{(2)a}_{n}$ above
should be regarded as the redefined functions in (\ref{redefinition}). Up to second order, the summed distribution functions
$\mathcal{J}^{I}_{n}$ and $\mathcal{J}^{a}_{n}$ satisfy the same equations as (\ref{cke-I-2-final}) and (\ref{cke-a-2-final}) except that
all the  superscripts $(0)$, $(1)$ and $(2)$ denoting the order indices  should be dropped.

\section{Summary}
\label{sec:summary}

Perturbation methods play a central role in the calculation involved with quantum field theory. A proper and consistent perturbation scheme makes it more
convenient and concise to solve some  theoretical problems order by order. The covariant gradient expansion is a powerful perturbation scheme to disentangle
the quantum transport equations from Abelian gauge field. However when we generalize this perturbation method to the non-Abelian case, some constraint conditions
 arise which is absent in Abelian case. We find that these constraint conditions originate from the fact that
 this expansion is not completely identical to the semiclassical expansion in powers of $\hbar$  for non-Abelian gauge field though it is identical for Abelian gauge field.
 We present a new perturbation scheme that makes the covariant gradient expansion and semiclassical expansion compatible with each other by rescaling the gauge potential
 and  field strength tensor.   By virtue of this new perturbation scheme, we derive the non-Abelian chiral kinetic equations with the Wigner function approach
 up  to  second order and find that some Wigner equations  hold automatically and  no longer lead to constraint equations any more.
 We integrate the covariant chiral kinetic equations  in eight-dimensional phase space  and obtain the chiral kinetic equations in  seven-dimensional phase space, which
 is more  suitable for  numerical evaluation.

\acknowledgments

This work was supported in part by  the National Natural
Science Foundation of China  under Grant
Nos. 12175123, 12321005

\appendix
\section{Some calculation details}

In this appendix, we present some calculation details  on the equations (\ref{cke-2-I-formal}) and (\ref{cke-2-a-formal}).
 The main task is to calculate the commutator relations.

In  the color singlet equations (\ref{cke-2-a-formal}), the last three terms associated with commutator
can be calculated  as follows,
\begin{eqnarray}
& &\nabla^{(1)Ib}_\mu{\dot x}^{(1)bc,\mu} -\dot{x}^{(1)I,\mu}\nabla_\mu^{(1)Ic} \nonumber\\
&=& \frac{1}{2N}  {F}_{\mu \nu}^b \bar\partial^\nu_p
\left( \frac{\lambda s}{2|\bar p|^2}\bar\epsilon^{\mu\alpha\beta}\bar p_\beta \mathscr{D}^{ bc}_\alpha\right)
- \frac{\lambda s}{2|\bar p|^2}\bar\epsilon^{\mu\alpha\beta}\bar p_\beta \partial^x_\alpha
\left(\frac{1}{2N}  {F}_{\mu \nu}^c \bar\partial^\nu_p\right)\nonumber\\
&=& \frac{\lambda s}{4 N |\bar p|^2}  {F}_{\mu \nu}^b
\left(\bar\epsilon^{\mu\alpha\nu} +\frac{2\bar p^\nu}{|\bar p|^2}\bar \epsilon^{\mu\alpha\beta}\bar p_\beta \right)\mathscr{D}^{ bc}_\alpha
- \frac{\lambda s}{4N|\bar p|^2}\bar\epsilon^{\mu\alpha\beta}\bar p_\beta
\left(\mathscr{D}_{\alpha}^{cb} {F}_{\mu \nu}^b \right)\bar\partial^\nu_p\nonumber\\
&=& \frac{\lambda s}{4 N |\bar p|^4}
\bar\epsilon^{\mu\nu\beta}\bar p_\beta {F}_{\mu \nu}^b \bar p^\alpha \mathscr{D}^{ bc}_\alpha
- \frac{\lambda s}{4N|\bar p|^2}\bar\epsilon^{\mu\alpha\beta}\bar p_\beta
\left(\mathscr{D}_{\alpha}^{cb} {F}_{\mu \nu}^b \right)\bar\partial^\nu_p
\end{eqnarray}
\begin{eqnarray}
& &{\mathscr{D}}^{ca}_\mu \dot{x}^{(2)Ia,\mu}\nonumber\\
&=& - \frac{ \lambda s}{8N|\bar p|^3} \bar\epsilon^{\nu\alpha\beta}\bar p_\beta \hat{\bar p}^\mu
\left({\mathscr{D}}^{ca}_\mu{{ F}^{a}_{\alpha\nu}}\right)
+ \frac{s }{4N|\bar p|^3}\bar\epsilon^{\mu\alpha\beta}\bar p_\beta \left( {\mathscr{D}}^{ca}_\mu {{ F}^{a}_{\alpha n}}\right)
\hspace{2cm} \nonumber\\
& & +  \bar\partial^\beta_p \left[\frac{\lambda  s }{4N|\bar p|^2}\bar\epsilon^{\mu\alpha\nu}\bar p_\nu
\left({\mathscr{D}}^{ca}_\mu { F}^{a}_{\alpha\beta} \right)\right]
\end{eqnarray}

\begin{eqnarray}
& &\left({\mathscr{D}}^{ca}_\mu\Omega^{(2)Ia} \right) \dot{x}^{(0)\mu}\nonumber\\
&=&-\frac{\lambda s}{8 N |\bar p|^3}\bar \epsilon^{\nu\alpha\beta}\bar p_\beta \left({\mathscr{D}}^{ca}_\mu F_{\alpha\nu}^a\right)  \dot{x}^{(0)\mu}\hspace{6cm}
\end{eqnarray}
Summing them gives rise to
\begin{eqnarray}
& &\nabla^{(1)Ib}_\mu{\dot x}^{(1)bc,\mu} -\dot{x}^{(1)I,\mu}\nabla_\mu^{(1)Ic}
+{\mathscr{D}}^{ca}_\mu \dot{x}^{(2)Ia,\mu} + \left({\mathscr{D}}^{ca}_\mu\sqrt{\omega}^{(2)Ia} \right) \dot{x}^{(0)\mu} \nonumber\\
&=&\frac{\lambda s}{4 N |\bar p|^2} \bar\epsilon^{\mu\nu\beta}\bar p_\beta {F}_{\mu \nu}^b \bar p^\alpha \mathscr{D}^{ bc}_\alpha
+ \frac{\lambda s}{4N|\bar p|^2}\bar\epsilon^{\mu\alpha\nu}\bar p_\nu
\left(\mathscr{D}_{\beta}^{ca} {F}_{\alpha\mu }^a \right)\bar\partial^\beta_p\nonumber\\
\end{eqnarray}
In  the color multiplet equations (\ref{cke-2-a-formal}), the last three terms associated with commutator
can be calculated  similarly,
\begin{eqnarray}
& &\nabla^{(1)a}_\mu \dot{x}^{(1)I,\mu} -\dot{x}^{(1)ab,\mu}{\nabla}^{(1) b}_\mu\nonumber\\
&=&  {F}_{\mu \nu}^a \bar\partial^\nu_p\left( \frac{\lambda s}{2|\bar p|^2}\bar\epsilon^{\mu\alpha\beta}\bar p_\beta \partial^x_\alpha\right)
- \frac{\lambda s}{2|\bar p|^2}\bar\epsilon^{\mu\alpha\beta}\bar p_\beta \mathscr{D}^{ ab}_\alpha  {F}_{\mu \nu}^b \bar\partial^\nu_p\nonumber\\
&=&\frac{\lambda s}{2|\bar p|^4}
\bar\epsilon^{\mu\nu\beta}\bar p_\beta \bar p^\alpha  {F}_{\mu \nu}^a \partial_\alpha^x
- \frac{\lambda s}{2|\bar p|^2}\bar\epsilon^{\mu\alpha\beta}\bar p_\beta\left( \mathscr{D}^{ ab}_\alpha  {F}_{\mu \nu}^b\right)  \bar\partial^\nu_p
\end{eqnarray}
\begin{eqnarray}
& &\mathscr{D}^{ab,\mu}\dot{ x}_\mu^{(2)b}\nonumber\\
&=& - \frac{ \lambda s}{4|\bar p|^3} \bar\epsilon^{\nu\alpha\beta}\bar p_\beta \hat{\bar p}^\mu
\left({\mathscr{D}}^{ab}_\mu{{ F}^{b}_{\alpha\nu}}\right)
+ \frac{s }{2|\bar p|^3}\bar\epsilon^{\mu\alpha\beta}\bar p_\beta \left( {\mathscr{D}}^{ab}_\mu {{ F}^{b}_{\alpha n}}\right)
 \nonumber\\
& & +  \bar\partial^\beta_p \left[\frac{\lambda  s }{2|\bar p|^2}\bar\epsilon^{\mu\alpha\nu}\bar p_\nu
\left({\mathscr{D}}^{ab}_\mu { F}^{b}_{\alpha\beta} \right)\right]
\end{eqnarray}

\begin{eqnarray}
& &\left(\mathscr{D}^{ab,\mu} \Omega^{(2)b } \right)\dot{ x}_\mu^{(0)} \nonumber\\
&=&-\frac{\lambda s}{4 |\bar p|^3}\bar \epsilon^{\nu\alpha\beta}\bar p_\beta \left({\mathscr{D}}^{ab}_\mu F_{\alpha\nu}^b\right)  \dot{x}^{(0)\mu}\hspace{4cm}
\end{eqnarray}
We can sum over these three terms and obtain
\begin{eqnarray}
& &\nabla^{(1)a}_\mu \dot{x}^{(1)I,\mu} -\dot{x}^{(1)ab,\mu}{\nabla}^{(1) b}_\mu
+{\mathscr{D}}^{ab}_\mu \dot{x}^{(2)b,\mu} + \left({\mathscr{D}}^{ab}_\mu\Omega^{(2)b} \right) \dot{x}^{(0)\mu} \nonumber\\
&=&\frac{\lambda s}{2 |\bar p|^4} \bar\epsilon^{\mu\nu\beta}\bar p_\beta {F}_{\mu \nu}^a \bar p^\alpha\partial^x_\alpha
+ \frac{\lambda s}{2 |\bar p|^2}\bar\epsilon^{\mu\alpha\nu}\bar p_\nu
\left(\mathscr{D}_{\beta}^{ab} {F}_{\alpha\mu }^b \right)\bar\partial^\beta_p
\end{eqnarray}
There are other  five   commutators in the chiral kinetic equation for the color multiplet.
The first term is given by
\begin{eqnarray}
\mathscr{D}^{ab}_\mu   {\dot x}^{(1)\mu,bc}-\dot{x}^{(1)\mu, ab}{\mathscr{D}}^{bc}_\mu
&=& \frac{\lambda s}{2|\bar p|^2} g f^{bca}\bar\epsilon^{\mu\alpha\beta}\bar p_\beta F^b_{\mu\alpha}
\end{eqnarray}
The remaining four terms are given by
\begin{eqnarray}
& &\nabla^{(1)a b}_\mu {\dot x}^{(1)bc,\mu}  -\dot{x}^{(1)ab,\mu}{\nabla}^{(1)bc}_\mu\nonumber\\
&=& \frac{1}{2}d^{bea}  {F}_{\mu \nu}^e \bar\partial^\nu_p
\left( \frac{\lambda s}{2|\bar p|^2}\bar\epsilon^{\mu\alpha\beta}\bar p_\beta \mathscr{D}^{ bc}_\alpha\right)
- \frac{\lambda s}{2|\bar p|^2}\bar\epsilon^{\mu\alpha\beta}\bar p_\beta \mathscr{D}^{ab}_\alpha
\left(\frac{1}{2}d^{ceb}  {F}_{\mu \nu}^e \bar\partial^\nu_p\right)\nonumber\\
&=& \frac{\lambda s}{4 |\bar p|^4}
\bar\epsilon^{\mu\nu\beta}\bar p_\beta  \bar p^\alpha
d^{bea}  {F}_{\mu \nu}^e\mathscr{D}^{ bc}_\alpha
+ \frac{\lambda s}{4|\bar p|^2}\bar\epsilon^{\mu\alpha\beta}\bar p_\beta
\left(d^{bea}{F}_{\mu \nu}^e \mathscr{D}_{\alpha}^{bc}  -\mathscr{D}_{\alpha}^{ab} d^{ceb}{F}_{\mu \nu}^e \right)
\bar\partial^\nu_p\nonumber\\
&=&  \frac{\lambda s}{4 |\bar p|^4}
\bar\epsilon^{\mu\nu\beta}\bar p_\beta  \bar p^\alpha
d^{bea}  {F}_{\mu \nu}^e\mathscr{D}^{ bc}_\alpha
-\frac{\lambda s}{4|\bar p|^2}\bar\epsilon^{\mu\alpha\beta}\bar p_\beta d^{cab}\left(\mathscr{D}_{\alpha}^{be}{F}_{\mu \nu}^e\right)\bar\partial^\nu_p
\end{eqnarray}

\begin{eqnarray}
& &\nabla_\mu^{(2)ac} \dot{x}^{(0)\mu}-\dot{x}^{(0)\mu} \nabla_\mu^{(2)ac}\nonumber\\
&=& -\frac{1}{8} f^{cba} \bar\partial^\nu_p \bar\partial^\lambda_p  \left(\mathscr{D}_\lambda^{be}{ F}^e_{\mu \nu}\right)  \dot{x}^{(0)\mu}
+\frac{1}{8} f^{cba} \dot{x}^{(0)\mu} \bar\partial^\nu_p \bar\partial^\lambda_p  \left(\mathscr{D}_\lambda^{be}{ F}^e_{\mu \nu}\right)
\nonumber\\
&=& -\frac{1}{8} f^{cba} \bar\partial^\nu_p \left( \frac{\bar p^\mu \bar p^\lambda -\bar p^2 \Delta^{\mu\lambda}}{|\bar p|^3}\right)  \left(\mathscr{D}_\lambda^{be}{ F}^e_{\mu \nu}\right)
 -\frac{1}{8} f^{cba}  \left( \frac{\bar p^\mu \bar p^\nu -\bar p^2 \Delta^{\mu\nu}}{|\bar p|^3}\right)
\bar\partial^\lambda_p   \left(\mathscr{D}_\lambda^{be}{ F}^e_{\mu \nu}\right)\nonumber\\
&=& -\frac{1}{8|\bar p|^3} f^{cba}  \left(\bar p^\mu \bar p^\nu -\bar p^2 \Delta^{\mu\nu}\right)
\left(\mathscr{D}_\nu^{be}{ F}^e_{\mu \lambda}\right) \bar\partial^\lambda_p
\end{eqnarray}

\begin{eqnarray}
& &\mathscr{D}^{ab,\mu}\dot{ x}_\mu^{(2)bc}
- \dot{x}^{(2)ab,\mu}{\mathscr{D}}^{bc}_\mu  \nonumber\\
&=& - \frac{ \lambda s}{8|\bar p|^3} \bar\epsilon_{\gamma\alpha\beta}\bar p^\beta
\left(\mathscr{D}^{ab,\mu} d^{ceb}{{ F}^{e,\alpha\gamma}} - d^{bea}{{ F}^{e,\alpha\gamma}}\mathscr{D}^{bc,\mu}  \right)\hat{\bar p}_\mu
\nonumber\\
& &+ \frac{s }{4|\bar p|^3}\bar\epsilon_{\mu\alpha\beta}\bar p^\beta
\left(\mathscr{D}^{ab,\mu} d^{ceb} {{ F}^{e,\alpha n}} -  d^{bea} {{ F}^{e,\alpha n}} \mathscr{D}^{bc,\mu}\right)\nonumber\\
& &+  \frac{\lambda}{8|\bar p|^3}\bar p^\alpha \left(\hat{\bar p}_\mu n^\nu -\lambda \Delta^\nu_\mu\right)
\left(\mathscr{D}^{ab,\mu}f^{ceb}{ F}_{\alpha \nu}^e - f^{bea}{ F}_{\alpha \nu}^e  \mathscr{D}^{bc,\mu}\right)
\nonumber\\
& &+  \frac{\lambda}{8|\bar p|^2}\Delta^\nu_\mu\left(\mathscr{D}^{ab,\mu} f^{ceb}{ F}_{\nu n}^e
-  f^{bea}{ F}_{\nu n}^e  \mathscr{D}^{bc,\mu}\right) \nonumber\\
& & - \frac{ 1 }{8|\bar p|^5}
\left[ \bar p_\alpha \bar p_\beta\bar p_\mu
+ \bar p^2\left( \Delta_{\alpha\beta}\bar p_\mu
-2\Delta_{\mu\beta}\bar p_\alpha \right) \right]
\left(\mathscr{D}^{ab,\mu}{\mathscr{D}}^{be,\alpha} {\mathscr{D}}^{ec,\beta}
-{\mathscr{D}}^{ae,\alpha} {\mathscr{D}}^{eb,\beta}\mathscr{D}^{bc,\mu} \right) \nonumber\\
& &  +  \bar\partial_\beta^p
\left[\frac{\lambda  s }{4|\bar p|^2}\bar\epsilon_{\mu\alpha\nu}\bar p^\nu
\left(\mathscr{D}^{ab,\mu} d^{ceb} { F}^{e,\alpha\beta}
- d^{bea} { F}^{e,\alpha\beta} \mathscr{D}^{bc,\mu} \right)\right.\nonumber\\
& &\hspace{1cm}\left.+\frac{1}{8|\bar p|^3}\left({\bar p}_\mu \bar p_\alpha -\bar p^2 \Delta_{\alpha\mu}\right)
\left(\mathscr{D}^{ab,\mu}f^{ceb} {F}^{e,\alpha \beta}
-f^{bea} {F}^{e,\alpha \beta} \mathscr{D}^{bc,\mu}\right) \right]
\end{eqnarray}

\begin{eqnarray}
&=&  \frac{ \lambda s}{8|\bar p|^3} \bar\epsilon_{\mu\alpha\beta}\bar p^\beta
\left(\mathscr{D}^{ab,\nu} d^{ceb}{{ F}^{e,\alpha\mu}} - d^{bea}{{ F}^{e,\alpha\mu}}\mathscr{D}^{bc,\nu}  \right)\hat{\bar p}_\nu
\nonumber\\
& &+ \frac{s }{4|\bar p|^3}\bar\epsilon_{\mu\alpha\beta}\bar p^\beta
\left(\mathscr{D}^{ab,\mu} d^{ceb} {{ F}^{e,\alpha \nu}} -  d^{bea} {{ F}^{e,\alpha \nu}} \mathscr{D}^{bc,\mu}\right) n_\nu\nonumber\\
& &+  \frac{\lambda}{8|\bar p|^3}\bar p^\alpha \left(\hat{\bar p}^\mu n^\nu -\lambda \Delta^{\nu\mu}\right)
\left(\mathscr{D}^{ab}_\mu f^{ceb}{ F}_{\alpha \nu}^e - f^{bea}{ F}_{\alpha \nu}^e  \mathscr{D}^{bc}_\mu\right)
\nonumber\\
& &+  \frac{\lambda}{8|\bar p|^2}\Delta^\nu_\mu\left(\mathscr{D}^{ab,\mu} f^{ceb}{ F}_{\nu n}^e
-  f^{bea}{ F}_{\nu n}^e  \mathscr{D}^{bc,\mu}\right) \nonumber\\
& & + \frac{ 1 }{8|\bar p|^3}
\left( \Delta_{\alpha\beta}\bar p_\mu -2\Delta_{\mu\beta}\bar p_\alpha \right)
\left(\mathscr{D}^{ab,\mu}{\mathscr{D}}^{be,\alpha} {\mathscr{D}}^{ec,\beta}
-{\mathscr{D}}^{ae,\alpha} {\mathscr{D}}^{eb,\beta}\mathscr{D}^{bc,\mu} \right) \nonumber\\
& &  +
\frac{\lambda  s }{4|\bar p|^2}\bar\epsilon_{\mu\alpha\beta}\bar p^\beta
\left(\mathscr{D}^{ab,\mu} d^{ceb} { F}^{e,\alpha\nu}
- d^{bea} { F}^{e,\alpha\nu} \mathscr{D}^{bc,\mu} \right)\bar\partial_\nu^p\nonumber\\
& &+\frac{1}{8|\bar p|^3}\left({\bar p}_\mu \bar p_\alpha -\bar p^2 \Delta_{\alpha\mu}\right)
\left(\mathscr{D}^{ab,\mu}f^{ceb} {F}^{e,\alpha \nu}
-f^{bea} {F}^{e,\alpha \nu} \mathscr{D}^{bc,\mu}\right) \bar\partial_\nu^p
\end{eqnarray}

\begin{eqnarray}
&=&  \frac{ \lambda s}{8|\bar p|^3} \bar\epsilon^{\mu\alpha\beta}\bar p_\beta \hat{\bar p}^\nu
d^{cba}\left(  \mathscr{D}^{be}_\nu {{ F}^{e}_{\alpha\mu}} \right)
+ \frac{s }{4|\bar p|^3}\bar\epsilon^{\mu\alpha\beta}\bar p_\beta n_\nu
d^{cba} \left( \mathscr{D}^{be}_\mu  { F}^{e}_{\alpha \nu} \right) \nonumber\\
& &+  \frac{\lambda}{8|\bar p|^3} \left( \bar p^\alpha \hat{\bar p}^\mu n^\nu -|\bar p| n^\alpha\Delta^{\mu\nu} -\lambda \bar p^\alpha \Delta^{\mu\nu}\right)
f^{cba}\left(\mathscr{D}^{be}_\mu { F}_{\alpha \nu}^e \right)
\nonumber\\
& & - \frac{ 1 }{8|\bar p|^3}
\left( \Delta^{\alpha\beta}\bar p^\mu -2\Delta^{\mu\beta}\bar p^\alpha \right)
\left( f^{bea}F^e_{\mu\alpha} {\mathscr{D}}^{bc}_{\beta}
+ {\mathscr{D}}^{ab}_{\alpha} f^{ceb}F^e_{\mu\beta}\right) \nonumber\\
& &  + \frac{\lambda  s }{4|\bar p|^2}\bar\epsilon^{\mu\alpha\beta}\bar p_\beta d^{cba}
\left(\mathscr{D}^{be}_\mu { F}^{e}_{\alpha\nu} \right) \bar\partial^\nu_p
+\frac{1}{8|\bar p|^3}\left({\bar p}^\mu \bar p^\alpha -\bar p^2 \Delta^{\alpha\mu}\right)
f^{cba}\left(\mathscr{D}^{be}_\mu {F}^e_{\alpha \nu}\right) \bar\partial^\nu_p
\end{eqnarray}

\begin{eqnarray}
&=&  \frac{ \lambda s}{8|\bar p|^3} \bar\epsilon^{\mu\alpha\beta}\bar p_\beta \dot{x}^{(0)\nu}
d^{cba}\left(  \mathscr{D}^{be}_\nu {{ F}^{e}_{\alpha\mu}} \right)
+  \frac{\lambda}{8|\bar p|^3} \left( \bar p^\alpha \hat{\bar p}^\mu n^\nu - \lambda |\bar p| \dot{x}^{(0)\alpha}\Delta^{\mu\nu} \right)
f^{cba}\left(\mathscr{D}^{be}_\mu { F}_{\alpha \nu}^e \right)
\nonumber\\
& & - \frac{ 1 }{8|\bar p|^3}
\Delta^{\alpha\beta}\bar p^\mu \left(3 f^{bea}F^e_{\mu\alpha} {\mathscr{D}}^{bc}_{\beta}
+ {\mathscr{D}}^{ab}_{\alpha} f^{ceb}F^e_{\mu\beta}\right) \nonumber\\
& &  + \frac{\lambda  s }{4|\bar p|^2}\bar\epsilon^{\mu\alpha\beta}\bar p_\beta d^{cba}
\left(\mathscr{D}^{be}_\mu { F}^{e}_{\alpha\nu} \right) \bar\partial^\nu_p
+\frac{1}{8|\bar p|^3}\left({\bar p}^\mu \bar p^\alpha -\bar p^2 \Delta^{\alpha\mu}\right)
f^{cba}\left(\mathscr{D}^{be}_\mu {F}^e_{\alpha \nu}\right) \bar\partial^\nu_p
\end{eqnarray}

\begin{eqnarray}
& &\mathscr{D}^{ab,\mu} \Omega^{(2)bc} \dot{ x}_\mu^{(0)}-\Omega^{(2)ab}\dot{x}^{(0)\mu}{\mathscr{D}}^{bc}_\mu \nonumber\\
&=& -\frac{\lambda s}{8|\bar p|^3} \bar\epsilon_{\nu\alpha\beta}\bar p^\beta
\left({\mathscr{D}}^{ab,\mu} d^{ceb}F^{e,\alpha\nu} \dot{x}^{(0)}_\mu
-  d^{bea}F^{e,\alpha\nu} \dot{x}^{(0)}_\mu{\mathscr{D}}^{bc,\mu} \right)\nonumber\\
& &-\frac{1}{4|\bar p|^4}\left(\bar p^\alpha \bar p^\beta - \bar p^2 \Delta^{\alpha\beta}\right)
\left({\mathscr{D}}^{ab}_\mu\bar{\mathscr{D}}^{be}_\alpha \bar{\mathscr{D}}^{ec}_\beta \dot{x}^{(0)\mu}
- \bar{\mathscr{D}}^{ae}_\alpha \bar{\mathscr{D}}^{eb}_\beta \dot{x}^{(0)\mu}{\mathscr{D}}^{bc}_\mu\right)
  \nonumber\\
& &-\frac{\lambda }{8|\bar p|^3} \left({\mathscr{D}}^{ab}_\mu f^{ceb}  F_{\nu n}^e \dot{x}^{(0)\mu}
- f^{bca}  F_{\nu n}^c \dot{x}^{(0)\mu}{\mathscr{D}}^{bc}_\mu\right) \bar p^\nu\nonumber\\
& &-\bar\partial^\nu_p \left[\frac{\lambda }{8|\bar p|}
\left({\mathscr{D}}^{ab}_\mu f^{ceb}  F_{n \nu}^e \dot{x}^{(0)\mu}
 -  f^{bca}  F_{n \nu}^c \dot{x}^{(0)\mu}{\mathscr{D}}^{bc}_\mu \right)  \right]
\end{eqnarray}

\begin{eqnarray}
&=& -\frac{\lambda s}{8|\bar p|^3} \bar\epsilon_{\mu\alpha\beta}\bar p^\beta
\left({\mathscr{D}}^{ab,\nu} d^{ceb}F^{e,\alpha\mu}
-  d^{bea}F^{e,\alpha\mu}{\mathscr{D}}^{bc,\nu} \right) \dot{x}^{(0)}_\nu\nonumber\\
& &-\frac{1}{4|\bar p|^4}\left(\bar p^\alpha \bar p^\beta - \bar p^2 \Delta^{\alpha\beta}\right)
\left(f^{eba}F^e_{\mu\alpha} \bar{\mathscr{D}}^{bc}_\beta
- \bar{\mathscr{D}}^{ab}_\alpha f^{ecb}F^e_{\beta\mu}  \right)\dot{x}^{(0)\mu}
  \nonumber\\
& &-\frac{\lambda }{8|\bar p|^3} \bar p^\nu \dot{x}^{(0)}_\mu n^\alpha \left({\mathscr{D}}^{ab,\mu} f^{ceb}  F_{\nu \alpha}^e
- f^{bea}  F_{\nu \alpha}^e{\mathscr{D}}^{bc,\mu}\right)\nonumber\\
& &- \frac{1}{8|\bar p|^3}\bar p^\nu n^\mu
\left({\mathscr{D}}^{ab}_\mu f^{ceb}  F_{n \nu}^e
 -  f^{bea}  F_{n \nu}^e {\mathscr{D}}^{bc}_\mu \right)\nonumber\\
& &- \frac{\lambda }{8|\bar p|^4}\left(2 \bar p^\mu \bar p^\nu -\bar p^2 \Delta^{\mu\nu}\right)
\left({\mathscr{D}}^{ab}_\mu f^{ceb}  F_{n \nu}^e
 -  f^{bea}  F_{n \nu}^e {\mathscr{D}}^{bc}_\mu \right) \nonumber\\
& &-\frac{\lambda }{8|\bar p|}
\left({\mathscr{D}}^{ab}_\mu f^{ceb}  F_{n \nu}^e \dot{x}^{(0)\mu}
 -  f^{bea}  F_{n \nu}^e \dot{x}^{(0)\mu}{\mathscr{D}}^{bc}_\mu \right) \bar\partial^\nu_p
\end{eqnarray}

\begin{eqnarray}
&=& -\frac{\lambda s}{8|\bar p|^3} \bar\epsilon^{\mu\alpha\beta}\bar p_\beta \dot{x}^{(0)\nu}
d^{cba}\left({\mathscr{D}}^{be}_\nu F^{e}_{\alpha\mu} \right) \nonumber\\
& &+\frac{1}{4|\bar p|^4}\left(\bar p^\alpha \bar p^\beta - \bar p^2 \Delta^{\alpha\beta}\right)\dot{x}^{(0)\mu}
\left(f^{bea}F^e_{\mu\alpha} \bar{\mathscr{D}}^{bc}_\beta
- \bar{\mathscr{D}}^{ab}_\alpha f^{ceb}F^e_{\beta\mu}  \right)
  \nonumber\\
& &- \frac{\lambda }{8|\bar p|^4}\left( \bar p^\mu \bar p^\nu -\bar p^2 \Delta^{\mu\nu}\right)
f^{cba} \left({\mathscr{D}}^{be}_\mu  F_{n \nu}^e \right)
-\frac{\lambda }{8|\bar p|}f^{cba}\dot{x}^{(0)\mu} \left({\mathscr{D}}^{be}_\mu  F_{n \nu}^e \right) \bar\partial^\nu_p
\end{eqnarray}
Summing these four terms gives rise to
\begin{eqnarray}
& &\nabla^{(1)a b}_\mu {\dot x}^{(1)bc,\mu}  -\dot{x}^{(1)ab,\mu}{\nabla}^{(1)bc}_\mu
+\nabla_\mu^{(2)ac} \dot{x}^{(0)\mu}-\dot{x}^{(0)\mu} \nabla_\mu^{(2)ac}\nonumber\\
& &+\mathscr{D}^{ab,\mu}\dot{ x}_\mu^{(2)bc}- \dot{x}^{(2)ab,\mu}{\mathscr{D}}^{bc}_\mu
+\mathscr{D}^{ab,\mu} \Omega^{(2)bc} \dot{ x}_\mu^{(0)}-\Omega^{(2)ab}\dot{x}^{(0)\mu}{\mathscr{D}}^{bc}_\mu\nonumber\\
&=& \frac{\lambda s}{4 |\bar p|^4} \bar\epsilon^{\mu\nu\beta}\bar p_\beta  \bar p^\alpha d^{bea}  {F}_{\mu \nu}^e\mathscr{D}^{ bc}_\alpha
{+\frac{\lambda}{2|\bar p|^4}\left(\bar p^\alpha \bar p^\beta - \bar p^2 \Delta^{\alpha\beta}\right)
f^{bea}F^e_{n \beta} \bar{\mathscr{D}}^{bc}_\alpha} \nonumber\\
& & + \frac{\lambda  s }{2|\bar p|^2}\bar\epsilon^{\mu\alpha\beta}\bar p_\beta d^{cba}
\left(\mathscr{D}^{be}_\mu { F}^{e}_{\alpha\nu} \right) \bar\partial^\nu_p
-\frac{\lambda }{8|\bar p|}f^{cba}\dot{x}^{(0)\mu} \left({\mathscr{D}}^{be}_\mu  F_{n \nu}^e \right) \bar\partial^\nu_p
\end{eqnarray}
In deriving the above results, we used the following  identities
\begin{eqnarray}
f^{ade}f^{bce}+ f^{bde}f^{cae} + f^{cde}f^{abe} &=& 0,\\
f^{ade}d^{bce}+ f^{bde}d^{cae} + f^{cde}d^{abe} &=& 0.
\end{eqnarray}
Substituting all these results into the equations (\ref{cke-2-I-formal}) and (\ref{cke-2-a-formal}) results in the final results (\ref{cke-I-2-final}) and (\ref{cke-a-2-final}),
respectively.




\end{document}